\documentclass[aps, prc, reprint, onecolumn, amsmath, amssymb, superscriptaddress, floatfix, nofootinbib, notitlepage]{revtex4-1} 

\usepackage[pdftex]{graphicx}
\DeclareGraphicsExtensions{.jpg,.png,.pdf}

\usepackage[utf8]{inputenc}
\usepackage{slashed}
\usepackage{hyperref}
\usepackage{amsmath}
\usepackage{amssymb}
\usepackage{amsfonts}
\usepackage{tabularx}
\usepackage{booktabs}
\usepackage{graphicx}
\usepackage{color}
\usepackage{multirow}
\usepackage[perpage,symbol]{footmisc}
\usepackage{verbatim}
\usepackage{dcolumn}
\usepackage{bm}
\usepackage[inline]{enumitem}
\definecolor{theblue}{RGB}{0,50,230}
\hypersetup{
	colorlinks=true,
	linkcolor=theblue,
	citecolor=theblue,
	urlcolor=theblue
}

\usepackage{lineno}

\begin{document}

\title{Perturbative and nonperturbative properties of heavy quark transport in a thermal SU(3) gluon plasma}
	
\author{Jiazhen~Peng}
\thanks{These authors contributed equally to this work.}%
\affiliation{%
	College of Mathematics and Physics, College of Nuclear Energy Science and Engineering, China Three Gorges University, Yichang 443002, China\\
}%
	
\author{Yage~Zhen}
\thanks{These authors contributed equally to this work.}%
\affiliation{%
	College of Mathematics and Physics, College of Nuclear Energy Science and Engineering, China Three Gorges University, Yichang 443002, China\\
}%

\author{Jiale~Lou}
\affiliation{%
	College of Mathematics and Physics, College of Nuclear Energy Science and Engineering, China Three Gorges University, Yichang 443002, China\\
}%
	
\author{Fei~Sun}
\affiliation{%
	College of Mathematics and Physics, College of Nuclear Energy Science and Engineering, China Three Gorges University, Yichang 443002, China\\
}%
\affiliation{%
	Center for Astronomy and Space Sciences, China Three Gorges University, Yichang 443002, China\\
}%
	
\author{Kejun~Wu}
\affiliation{%
	College of Mathematics and Physics, College of Nuclear Energy Science and Engineering, China Three Gorges University, Yichang 443002, China\\
}%
\affiliation{%
	Center for Astronomy and Space Sciences, China Three Gorges University, Yichang 443002, China\\
}%
	
\author{Wei~Xie}
\affiliation{%
	College of Mathematics and Physics, College of Nuclear Energy Science and Engineering, China Three Gorges University, Yichang 443002, China\\
}%
\affiliation{%
	Center for Astronomy and Space Sciences, China Three Gorges University, Yichang 443002, China\\
}%
	
\author{Zuman~Zhang}
\affiliation{%
	School of Physics and Mechanical Electrical and Engineering, Hubei University of Education, Wuhan 430205, China\\
}%
\affiliation{%
	Institute of Theoretical Physics, Hubei University of Education, Wuhan 430205, China\\
}%

\author{Shuang~Li}
\email[Contact author: ]{lish@ctgu.edu.cn}
\affiliation{%
	College of Mathematics and Physics, College of Nuclear Energy Science and Engineering, China Three Gorges University, Yichang 443002, China\\
}%
\affiliation{%
	Center for Astronomy 
	and Space Sciences, China Three Gorges University, Yichang 443002, China\\
}%
	
\author{Sa~Wang}
\email[Contact author: ]{wangsa@ctgu.edu.cn}
\affiliation{%
	College of Mathematics and Physics, College of Nuclear Energy Science and Engineering, China Three Gorges University, Yichang 443002, China\\
}%
\affiliation{%
	Center for Astronomy and Space Sciences, China Three Gorges University, Yichang 443002, China\\
}%
	
\date{\today}

\begin{abstract}
We investigate the perturbative and nonperturbative aspects of heavy quark transport in a thermal SU(3) gluon plasma.
Based on the soft-hard factorized model, we extend the original perturbative framework to the near-critical temperature region, where nonperturbative effects become significant.
The transition behavior of the semi-quark-gluon-plasma (semi-QGP) is described via a temperature-dependent background field incorporated in the background field effective theory.
By implementing this approach, we quantitatively evaluate the collisional energy loss and momentum diffusion coefficients of charm and bottom quarks as functions of the incoming energy and medium temperature.
Our results show a distinct suppression of both the energy loss and the diffusion coefficients relative to conventional perturbative estimates, especially near the critical temperature.
This suppression originates from the emergence of a temperature-dependent color background field, which effectively reduces the color charge screening of the medium.
These findings provide important theoretical insight into the phenomenology of heavy-flavor probes, offering a unified theoretical framework applicable across both high- and low-momentum regimes.
\end{abstract}
	
\maketitle
	
	
\section{INTRODUCTION}\label{sec:Intro}
The study of heavy quark transport in a deconfined medium has become a cornerstone for understanding the microscopic structure of the quark-gluon plasma (QGP) formed in high-energy nuclear collisions at RHIC and the LHC.
Due to their large masses ($m_Q \gg T$), heavy quarks serve as excellent probes of the medium’s dynamical properties.
The transport of these heavy quarks encodes critical information about the coupling strength, color screening, and near-critical dynamics of gauge matter.
Within the perturbative framework, the hard thermal loop (HTL) resummation and the next-to-leading order (NLO) kinetic calculations have provided systematic estimates of the heavy-quark drag and diffusion coefficients~\cite{Moore:2004tg, HAQUE2025104136, CaronHuot:2008uh}.
These approaches, however, are valid primarily in the weak-coupling limit at high temperatures ($T \gg T_c$).
In contrast, lattice-QCD~\cite{Banerjee:2012ra, Ding:2012sp, Altenkort:2023eav}, chromo-magnetic-monopole~\cite{Liao:2006ry, Liao:2008jg, Liao:2008dk, Xu:2015bbz} and effective $T$-matrix~\cite{vanHees:2007me, Riek:2010fk, He:2012df} studies have demonstrated that near the deconfinement transition, the strongly interacting plasma exhibits strong correlations that cannot be captured perturbatively.
Moreover, holographic models based on the AdS/CFT correspondence~\cite{Herzog:2006gh, Gubser:2006bz, Chen:2024epd, Chen:2025fpd} predict significantly larger drag coefficients, highlighting the importance of strong-coupling dynamics.

Despite the remarkable progress of both perturbative and nonperturbative approaches, a unified theoretical description that connects the high-energy perturbative regime with the near-critical strongly coupled regime remains elusive.
To address this gap, we extend the recently developed soft-hard factorized model (SHFM)~\cite{Peng:2024zvf, Li:2021nim, Lou:2025wmw} to incorporate a temperature-dependent background field, effectively bridging the perturbative and nonperturbative sectors.
In this framework, the soft medium excitations are dynamically dressed by the background field, while the hard scattering processes retain their perturbative character.
This formulation allows us to describe heavy quark transport consistently from the perturbative domain ($T\gg T_c$) down to the semi-QGP regime ($T\lesssim2-3T_c$).
A complete QCD analysis would require a nontrivial extension of the effective theory to include nonperturbative contributions from both thermal bosons and fermions.
As a first step, this work considers only the bosonic contributions, i.e., an SU(3) gluon plasma, leaving the full fermionic sector for future study.

Recent experimental measurements of heavy-flavor observables at RHIC and LHC, including nuclear modification factors and elliptic flow coefficients~\cite{Dong:2019byy, Tang:2020ame, Chen:2024aom, ALICE:2021rxa, ALICE:2022wpn}, have further emphasized the necessity of such a unified framework.
The simultaneous description of these observables requires a theoretical model that captures both perturbative scattering at high energies and nonperturbative color correlations near the phase transition.
The present work aims to provide such a framework through a background-field-based extension of SHFM.

The four-momentum is denoted as capital letters $P^{\mu}=(p_{0},\vec{p}\;)$, while the three-momentum vector is represented as lowercase letters $\vec{p}$.
The spacetime vector indices are denoted by Greek symbols, $\mu,\nu\dots\in\{0,1,2,3\}$, while the color indices are represented by Latin ones, $i,j,\dots\in\{1,2,3\}$.
We work with natural units $\hbar=c=1$ and metric $g^{\mu\nu}=\rm {diag}(1,-1,-1,-1)$ throughout this manuscript.

This manuscript is organized as follows: In Sec.~\ref{sec:formalism}, we detail the theoretical framework with and without the background field, including the SHFM, the calculation of the HQ self-energy, and the derivation of the transport coefficients.
Section~\ref{sec:result} is devoted to presenting and discussing our main numerical results, with a focus on the influence of the background field on the energy loss and momentum diffusion coefficients.
Finally, we summarize our findings and provide an outlook in Section~\ref{sec:summary}.


\section{Framework and setup}\label{sec:formalism}

\subsection{Heavy quark transport in the soft-hard factorized approach}\label{subsec:SoftHardModel}
\begin{figure}[!htbp]
	\centering
	\includegraphics[width=.32\textwidth]{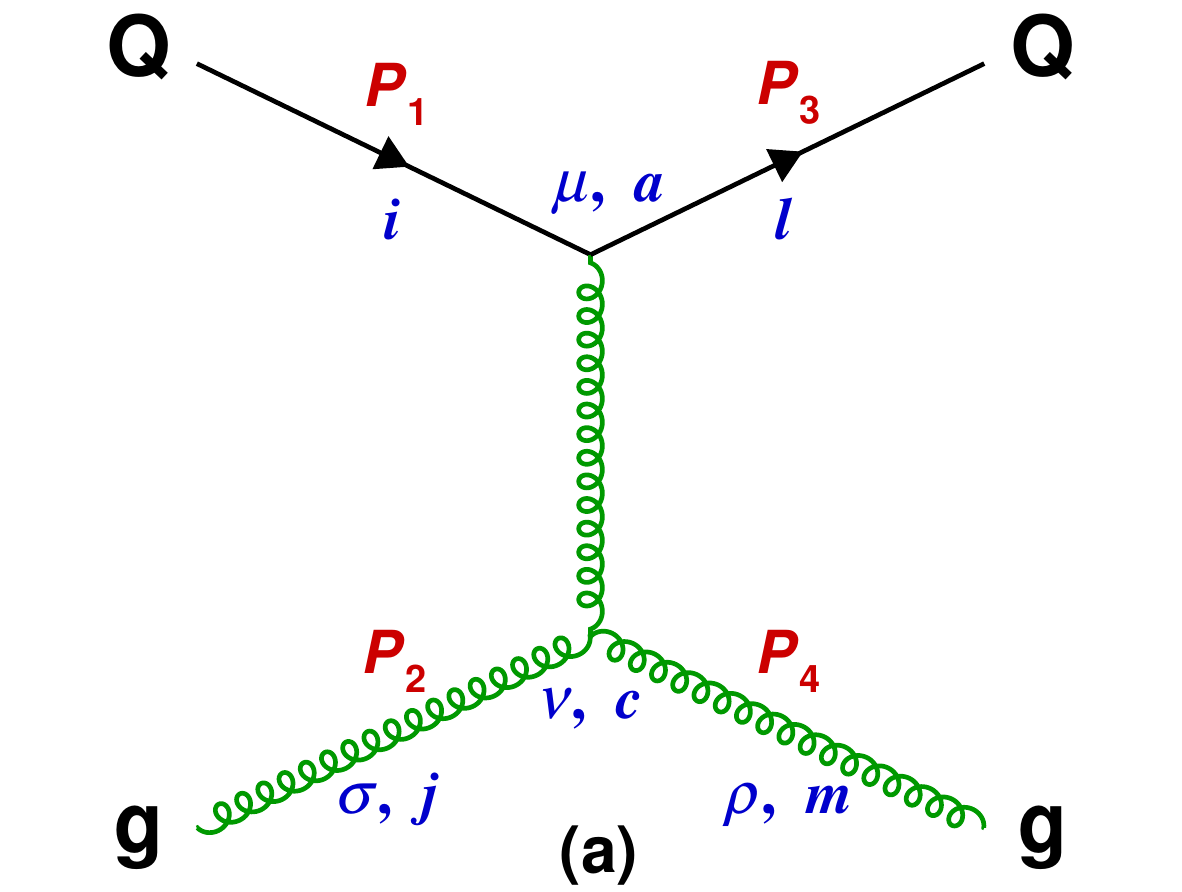}
	\includegraphics[width=.32\textwidth]{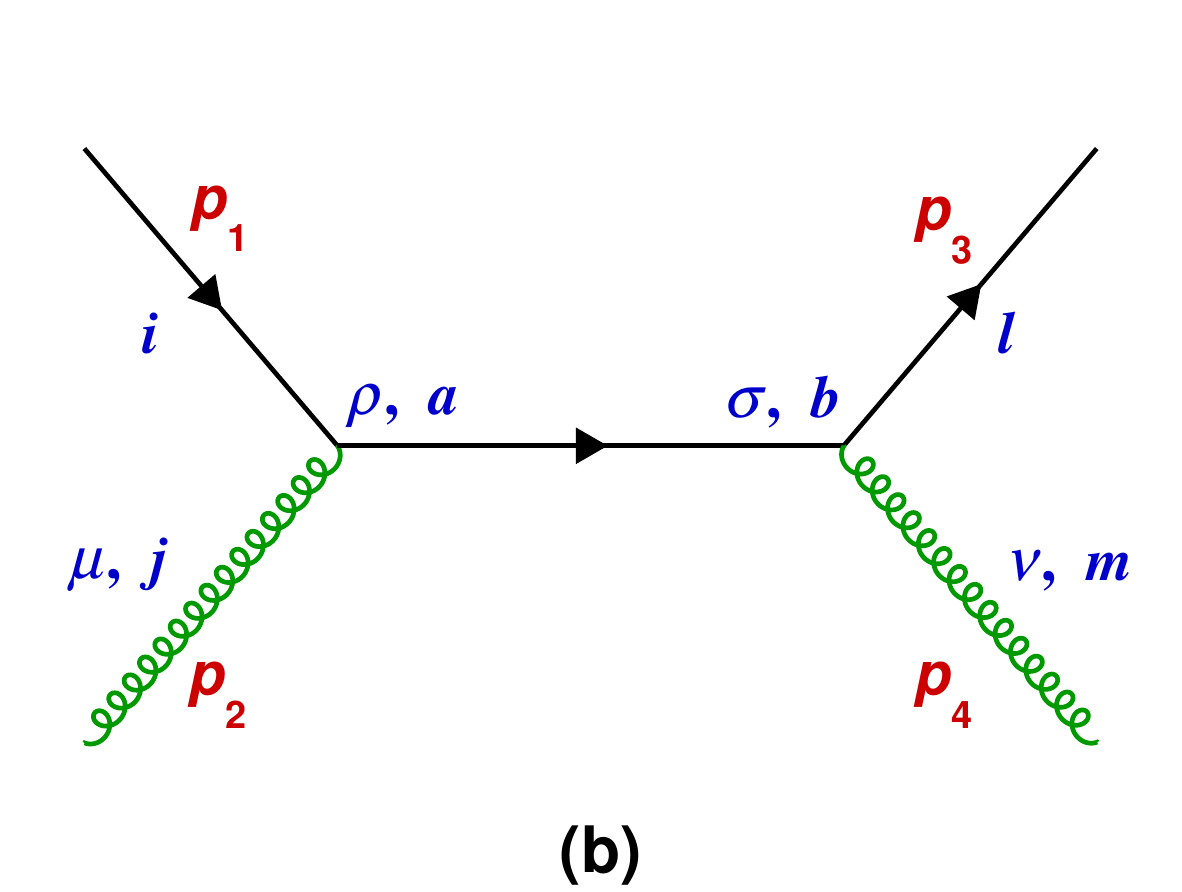}
	\includegraphics[width=.32\textwidth]{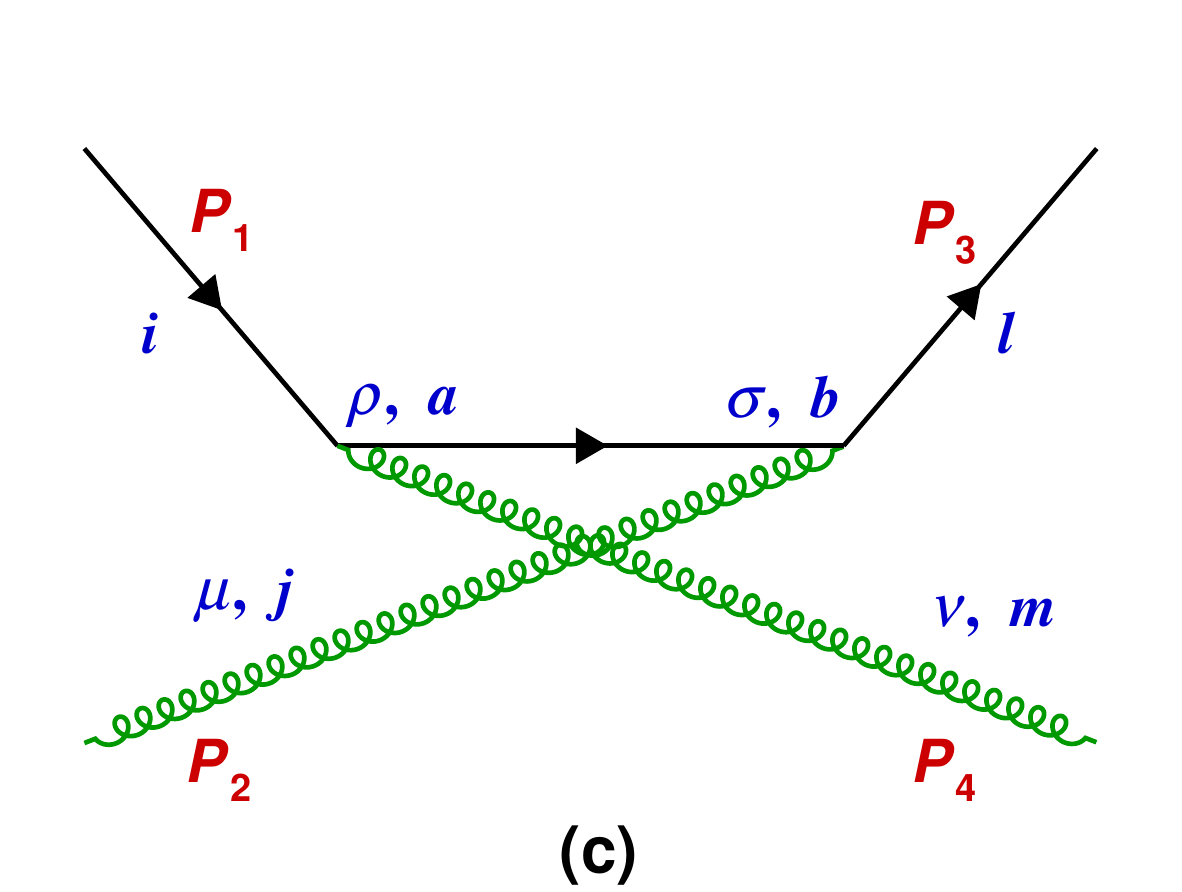}
	\caption{Tree-level Feynman diagrams for the scattering processes $Qg\to Qg$
		in $t$ (panel a), $s$ (panel b) and $u$ (panel c) channels in vacuum.}
	\label{fig:TwoBody_Diag}
\end{figure}

When propagating through a hot SU(3) gluon plasma consisting of thermal gluons,
a heavy quark ($Q$) undergoes scattering,
which can be characterized as the two-body elementary processes,
$Q~(P_{1})+g~(P_{2})\to Q~(P_{3})+g~(P_{4})$,
where $P_{1}=(E_{1},\vec{p}_{1})$ and $P_{3}=(E_{3},\vec{p}_{3})$
are the four-momentum of the injected heavy quark
before and after scattering. 
$P_{2}$ and $P_{4}$ are the ones for thermal gluon.
The corresponding tree-level Feynman diagrams for the $t$-, $s$- and $u$ channels
are presented in the panels a, b, c of Fig.~\ref{fig:TwoBody_Diag}, respectively.
For each diagram in Fig.~\ref{fig:TwoBody_Diag},
the corresponding matrix elements at leading order in $g$
can be found in Refs.~\cite{Peng:2024zvf, Li:2021nim},
and the four-momentum transfer is
$P^{\mu}_{1}-P^{\mu}_{3} = (\omega,~\vec{q}^{\;}) = (\omega,~\vec{q}_{T},~q_{L})$.
The related Mandelstam invariants, $t$, $s$ and $u$, are defined accordingly.

For small momentum transfers $t\to0$,
the $t$ channel gluon propagator (panel a in Fig.~\ref{fig:TwoBody_Diag})
causes an infrared divergence in the production cross-section.
Such divergence can be cured by taking into account the contributions from the long-wavelength gluons,
which correspond to small momentum transfer $\sqrt{-t}\sim gT$, i.e. soft scattering, in a thermal perturbation theory.
The soft gluon exchange in $t$ channel (see panel a in Fig.~\ref{fig:TwoBody_Diag})
features long-range interactions, and they are therefore screened by the medium partons.
Formally, the associated gluon propagator must be screened
with its self-energy~\cite{Braaten91PRL, JeanPR02}.
The contributions from the large momentum transfer $\sqrt{-t}\gtrsim T$, i.e. hard scattering,
where the Born approximation is valid, is straightforward to perform a pQCD calculation in this regime.
This is the soft-hard factorized approach~\cite{PhysRevD.44.1298, PhysRevD.44.R2625, Romatschke:2003vc, Romatschke:2004au, Djordjevic:2006tw, Peigne:2007sd, Peigne:2008nd, Alberico:2011zy},
which allows one to decompose the soft HQ-medium interactions with $-t<-t^{\ast}$, from the hard ones with $-t>-t^{\ast}$.
The intermediate scale $t^{\ast}$ is formally chosen as $m_{D}^{2}\ll -t^{\ast}\ll T^{2}$,
implying the weak-coupling or high-temperature limit~\cite{Braaten91PRL, Romatschke:2004au, Gossiaux:2008jv}.
The cutoff $|t^{\ast}|$ is usually characterized in terms of the gluon Debye screening mass squared $m_{D}^{2}$.

\subsubsection{The interaction rate for two-body scattering}\label{subsubsec:gamma_HTLpQCD}
With the soft-hard factorization model,
the scattering rate arises from three kinematic regimes:
\begin{enumerate}
	\item[(1)] Soft region ($-t < -t^{\ast}$):
	heavy quark scattering off medium partons via $t$ channel exchange with small momentum transfer;
	the interference with $s/u$ channels is neglected;
	\item[(2)] Hard region ($-t > -t^{\ast}$):
	heavy quark scattering off medium partons via $t$ channel exchange with large momentum transfer;
	the interference with $s/u$ channels is included and attributed to the $t$ channel;
	\item[(3)] $su$ region:
	heavy quarks scattering off thermal gluons via $s$- and $u$ channels for both small and large momentum transfers;
	this contribution is generally small compared to the $t$ channel.
\end{enumerate}
The total scattering rate is the sum of these three kinematic regions
\begin{equation}
	\begin{aligned}\label{eq:Gamma_Full_Def}
		\Gamma = \Gamma^{soft}_{(t)} + \Gamma^{hard}_{(t)} + \Gamma_{(su)},
	\end{aligned}
\end{equation}
in which the contributions from the $t$ channel are shown as the first ($soft$) and second ($hard$) terms on the right-hand side,
together with the contributions from $s$- and $u$ channels expressed as the third term.

\subsubsection{The collisional energy loss and transport coefficients of heavy quarks}\label{susubbsec:ElossCoefficients_Def}
The heavy quark energy loss per distance traveled is written as
\begin{equation}\label{eq:ELoss_Def}
	\begin{aligned}
		-\frac{dE}{dz}=\int d^{3}\vec{q} \; \frac{d\Gamma}{d^{3}\vec{q}} \; \frac{\omega}{v_{1}},
	\end{aligned}
\end{equation}
where, $v_{1}$ is the velocity of the heavy quark,
$d\Gamma/d^{3}\vec{q}$ is the differential scattering rate with respect to the three-momentum transfer $\vec{q}$
and the energy transfer is $\omega$.
The total energy loss in the soft-hard factorized approach is given by
inserting Eq.~(\ref{eq:Gamma_Full_Def}) into Eq.~(\ref{eq:ELoss_Def}), yielding
\begin{equation}
	\begin{aligned}\label{eq:dEdz_Full_Def}
		-\frac{dE}{dz} = \biggr[-\frac{dE}{dz}\biggr]^{soft}_{(t)} + \biggr[-\frac{dE}{dz}\biggr]^{hard}_{(t)} + \biggr[-\frac{dE}{dz}\biggr]_{(su)}.
	\end{aligned}
\end{equation}

It was argued~\cite{Moore:2004tg} that the interactions between HQs and the medium partons
can be encoded into the momentum diffusion coefficients,
\begin{equation}
	\begin{aligned}\label{eq:KappaT_Def}
		\kappa_{T} &= \frac{1}{2}  \int d^{3}\vec{q} \; \frac{d\Gamma}{d^{3}\vec{q}} \; \vec{q}^{\;2}_{T}= \frac{1}{2}  \int d^{3}\vec{q} \; \frac{d\Gamma}{d^{3}\vec{q}}\biggr[ \omega^{2} - t - \frac{(2\omega E_{1}-t)^{2}}{4\vec{p}_{1}^{\;2}} \biggr],
	\end{aligned}
\end{equation}
\begin{equation}
	\begin{aligned}\label{eq:KappaL_Def}
		\kappa_{L} &= \int d^{3}\vec{q} \frac{d\Gamma}{d^{3}\vec{q}} \; q^{2}_{L}
		= \frac{1}{4\vec{p}_{1}^{\;2}} \int d^{3}\vec{q} \frac{d\Gamma}{d^{3}\vec{q}} \; (2\omega E_{1}-t)^{2},
	\end{aligned}
\end{equation}
which quantifies the momentum fluctuations in the direction that is parallel                                                        
(i.e. longitudinal) and perpendicular (i.e. transverse) to the propagation, respectively.
In the framework of Langevin transport~\cite{HQQGPRapp10},
the drag coefficient, which describes the momentum/energy loss,
can be obtained via the dissipation-fluctuation relation
$\eta_{D}=\eta_{D}(\kappa_{T}, \kappa_{L})$.
See Ref.~\cite{Li:2019lex} for more details.
Similar with Eqs.~(\ref{eq:Gamma_Full_Def}) and (\ref{eq:dEdz_Full_Def}),
the complete results for the above coefficients can be formulated accordingly.

\subsection{Perturbative calculation of the collisional energy loss and transport coefficients in pure gauge theory}\label{subsec:PerturbativeModel}
According to Eqs.~(\ref{eq:ELoss_Def}),~(\ref{eq:KappaT_Def}) and (\ref{eq:KappaL_Def}),
one can see that the scattering rate [Eq.~(\ref{eq:Gamma_Full_Def})] is the key variable in this work.
In this Subsection, we calculate the scattering rate with the soft-hard factorization model,
including the thermal perturbative description of soft scatterings ($-t<-t^{\ast}$) and
a perturbative QCD-based calculation for hard collisions ($-t>-t^{\ast}$).
For the sake of completeness, we show the typical results with necessary explanations below.
The details are relegated to Refs.~\cite{Peng:2024zvf, Li:2021nim} and the references therein.

\subsubsection{The results in soft region $-t<-t^{\ast}$}\label{susubbsec:Pert_Soft}
It is argued~\cite{WeldonPRD83, PhysRevD.44.1298} that,
a heavy fermion propagating through a hot non-Abelian gauge plasma,
the scattering rate can be obtained from the imaginary part of the particle's self-energy $Im\Sigma$,
\begin{equation}
	\begin{aligned}\label{eq:Gamma_Soft1}
		\Gamma^{soft}_{(t)}(E_{1},T) &= -\frac{1}{2E_{1}} \bar{n}_{F}(E_{1}) Tr\bigr[ (\slashed{P}_{1} + m_{1})\cdot{Im}\Sigma (E_{1}+i\epsilon,\vec{p}_{1}) \bigr],
	\end{aligned}
\end{equation}
where, $m_{1}$ is the mass of the injected fermion.
For the heavy quark scattering off thermal gluons,
the heavy quark self-energy reads
\begin{equation}
	\begin{aligned}\label{eq:Sigma_Soft1}
		{\Sigma}(P_{1}) &= iC_{F}g^{2} \int\frac{d^{4}Q}{(2\pi)^{4}} D_{\mu\nu}(Q) \gamma^{\mu} \frac{1}{\slashed{P}_{1}-\slashed{Q}-m_{1}} \gamma^{\nu} \\
		&= -C_{F}g^{2}T \sum_{q_{0}}^{\;} \int\frac{d^{3}\vec{q}}{(2\pi)^{3}} D_{\mu\nu}(Q) \gamma^{\mu} \frac{1}{\slashed{P}_{1}-\slashed{Q}-m_{1}} \gamma^{\nu} ,
	\end{aligned}
\end{equation}
in Minkowski space. $Q=(q_{0},\vec{q}\;)$ is the four-momentum transfer here.
We make the replacement $\int{dq_{0}}/{(2\pi)} \to iT\sum_{q_{0}}$
in the second equality in Eq.~(\ref{eq:Sigma_Soft1}),
where the bosonic Matsubara frequency sum is over
the discrete imaginary values $q_{0}=i\omega_{n}=i2n\pi T$ ($n\in\mathcal{Z}$) for the gluon energy.
The transverse and longitudinal components of
the HTL-resummed gluon propagator $D_{\mu\nu}$, in Coulomb gauge,
is as follows~\cite{JeanPR02}
\begin{subequations}
	\begin{align}
		\Delta_{T}(q_{0},\vec{q}\;) &= \frac{-1}{q_{0}^{2}-\vec{q}^{\;2} - m_{D}^{2}\Pi_{T}} \label{eq:Propagator_Soft_T}, \\
		\Delta_{L}(q_{0},\vec{q}\;) &= \frac{-1}{\vec{q}^{\;2} + m_{D}^{2}\Pi_{L}} \label{eq:Propagator_Soft_L},
	\end{align}
\end{subequations}
where, $m_{D}^{2}$ is the Debye screening mass squared for gluon
\begin{equation}
	\begin{aligned}\label{eq:DebMas}
		m_{D}^{2}=\frac{N_{c}}{3} g^{2} T^{2}.
	\end{aligned}
\end{equation}
The transverse and longitudinal soft gluon self-energy are expressed as
\begin{subequations}
	\begin{align}
		{\Pi}_{T}(x) &= \frac{1}{2} \biggr[ x^{2}+(1-x^{2})\frac{x}{2}ln\frac{x+1}{x-1} \biggr] \label{eq:Pi_T}, \\
		{\Pi}_{L}(x) &=  1 - \frac{x}{2}ln\frac{x+1}{x-1} \label{eq:Pi_L},
	\end{align}
\end{subequations}
with $x=q_{0}/q$ and $q=|\vec{q}\;|$.

Then, we can calculate the mixed representation of $\Delta_{T/L}$ and
substitute it into Eq.~(\ref{eq:Sigma_Soft1}) and then into Eq.~(\ref{eq:Gamma_Soft1}) sequentially.
It yields~\cite{PhysRevD.44.1298, Li:2021nim}
\begin{equation}                                              \begin{aligned}\label{eq:Trace_Soft7}
		\Gamma^{soft}_{(t)}(E_{1},T) &= C_{F}g^{2} \int_q \int d\omega \;
		\delta(\omega-\vec{v}_{1}\cdot\vec{q}\;) \biggr\{ \mathcal{A}_{L} + \vec{v}_{1}^{\;2} \bigr[ 1-(\widehat{v}_{1}\cdot\widehat{q})^{2} \bigr]\mathcal{A}_{T} \biggr\},
	\end{aligned}
\end{equation}
with
\begin{equation}
	\begin{aligned} \label{eq:Pert_SH1}
		\mathcal{A}_{T/L}(\omega,q) &= \bar{n}_{F}(\omega)\cdot \rho_{T/L} (\omega,q).
	\end{aligned}
\end{equation}
$C_{F}=(N_{c}^{2}-1)/(2N_{c})$ is the quark Casimir factor for the fundamental representation
and $N_{c}$ is the color factor; $g^{2}=4\pi\alpha_{s}$ with the coupling constant $\alpha_{s}$;
$\vec{v}_{1}=\vec{p}_{1}/E_{1}$ denotes the HQ velocity.
The short-hand notation
\begin{equation}
	\begin{aligned}\label{eq:ShortNot_IntQ}
		\int_{q}\cdots =& \int\frac{d^{3}\vec{q}}{(2\pi)^{3}}\cdots,
	\end{aligned}
\end{equation}
denotes the phase space integrals.
$n_{B/F}$ indicates the thermal distributions for bosons/fermions,
\begin{equation}
	\begin{aligned}\label{eq:ThermalDis_Boson}
		n_{B}(E,T) &= \frac{1}{e^{E/T}-1},
	\end{aligned}
\end{equation}
\begin{equation}
	\begin{aligned}\label{eq:ThermalDis_Fermi}
		n_{F}(E,T) &= \frac{1}{e^{E/T}+1},
	\end{aligned}
\end{equation}
and
\begin{equation}
	\begin{aligned}\label{eq:ThermalDis_StatEff}
		\bar{n}_{B/F} = 1 \pm n_{B/F},
	\end{aligned}
\end{equation}
accounts for the Bose-enhancement or Pauli-blocking effect.
The transverse and longitudinal spectral density functions, as shown in Eq.~(\ref{eq:Trace_Soft7}),
are given by the imaginary part of the retarded propagator
\begin{equation}
	\begin{aligned}\label{eq:GammaRhoTL1}
		\rho_{T/L}(\omega,q) = 2\cdot Im\Delta_{T/L}(\omega+i0^{+},\vec{q}\;).
	\end{aligned}
\end{equation}
Inserting Eqs.~(\ref{eq:Propagator_Soft_T}) and (\ref{eq:Propagator_Soft_L}) into Eq.~(\ref{eq:GammaRhoTL1}), we have
\begin{subequations}
	\begin{align}
		\rho_{T}(\omega,q) = &\frac{\pi \omega m_{D}^{2}}{2q^{3}} (q^{2}-\omega^{2}) \biggr\{ \biggr[ q^{2}-\omega^{2} + \frac{\omega^{2}m_{D}^{2}}{2q^{2}} 
		\bigr(1+\frac{q^{2}-\omega^{2}}{2\omega q} ln\frac{q+\omega}{q-\omega}\bigr) \biggr]^{2} 
		+ \biggr[ \frac{\pi \omega m_{D}^{2}}{4q^{3}} (q^{2}-\omega^{2}) \biggr]^{2} \biggr\}^{-1}, \label{eq:RhoT_Soft_vsOmegaQ} \\
		\rho_{L}(\omega,q) = &\frac{\pi \omega m_{D}^{2}}{q}
		\biggr\{ \biggr[ q^{2}+m_{D}^{2}\bigr(1-\frac{\omega}{2q} ln\frac{q+\omega}{q-\omega}\bigr) \biggr]^{2} 
		+ \biggr( \frac{\pi \omega m_{D}^{2}}{2q} \biggr)^{2} \biggr\}^{-1}. \label{eq:RhoL_Soft_vsOmegaQ}
	\end{align}
\end{subequations}

Subsituting Eq.~(\ref{eq:Trace_Soft7}) back into Eqs.~(\ref{eq:ELoss_Def}), (\ref{eq:KappaT_Def}) and (\ref{eq:KappaL_Def}), respectively,
we get the energy loss, transverse and longitudinal momentum diffusion coefficients in soft collisions~\cite{Peng:2024zvf, Li:2021nim},
\begin{equation}
	\begin{aligned}\label{eq:dEdz_Soft_Perturbative}
		\biggr[-\frac{dE}{dz}\biggr]^{soft}_{(t)} =& \frac{C_{F}g^{2}}{8\pi^{2}v^{2}_{1}} \int^{0}_{t^{*}} dt \; (-t) \int_{0}^{v_{1}}dx \frac{x}{(1-x^{2})^{2}}
		\bigr[ \widetilde{\mathcal{A}}_{L-} + (v_{1}^{2}-x^{2}) \widetilde{\mathcal{A}}_{T-} \bigr],
	\end{aligned}
\end{equation}
\begin{equation}
	\begin{aligned}\label{eq:KappaT_Soft_Perturbative}
		\biggr[\kappa_{T}\biggr]^{soft}_{(t)} =& \frac{C_{F}g^{2}}{16\pi^{2}v_{1}^{3}} \int^{0}_{t^{\ast}} dt \; (-t)^{3/2}
		\int_{0}^{v_{1}}dx \frac{v_{1}^{2}-x^{2}}{(1-x^{2})^{5/2}}
		\bigr[ \widetilde{\mathcal{A}}_{L+} + (v_{1}^{2}-x^{2}) \widetilde{\mathcal{A}}_{T+} \bigr],
	\end{aligned}
\end{equation}
and
\begin{equation}
	\begin{aligned}\label{eq:KappaL_Soft_Perturbative}
		\biggr[\kappa_{L}\biggr]^{soft}_{(t)} =& \frac{C_{F}g^{2}}{8\pi^{2}v_{1}^{3}} \int^{0}_{t^{\ast}} dt \; (-t)^{3/2}
		\int_{0}^{v_{1}}dx \frac{x^{2}}{(1-x^{2})^{5/2}}
		\bigr[ \widetilde{\mathcal{A}}_{L+} + (v_{1}^{2}-x^{2}) \widetilde{\mathcal{A}}_{T+} \bigr].
	\end{aligned}
\end{equation}
with $t=\omega^2-q^{2}$ and $x=\omega/q$.
The factor $\widetilde{\mathcal{A}}_{T/L\pm}$ reads
\begin{equation}
	\begin{aligned} \label{eq:Pert_SH2}
		\widetilde{\mathcal{A}}_{T/L\pm} &= \mathcal{A}_{T/L}(\omega) \pm \mathcal{A}_{T/L}(-\omega).
	\end{aligned}
\end{equation}

To facilitate comparison with the results obtained in the presence of a background field
[Eqs.~(\ref{eq:dEdz_Soft_BFET}), (\ref{eq:KappaT_Soft_BFET}) and (\ref{eq:KappaL_Soft_BFET})],
we present the Eqs.~(\ref{eq:dEdz_Soft_Perturbative}), (\ref{eq:KappaT_Soft_Perturbative})
and (\ref{eq:KappaL_Soft_Perturbative}) in their current forms,
which can be further rewritten with
\begin{subequations}
	\begin{align}
		\widetilde{\mathcal{A}}_{T/L-} &= \rho_{T/L}, \\
		\widetilde{\mathcal{A}}_{T/L+} &= coth\left(\frac{\omega}{2T}\right) \cdot \rho_{T/L},
	\end{align}
\end{subequations}

\subsubsection{The results in hard region $-t>-t^{\ast}$}\label{susubbsec:Pert_Hard}
In hard collisions, as shown in Fig.~\ref{fig:TwoBody_Diag},
the heavy quark transition rate is defined as the rate of collisions with thermal gluon,
which changes the heavy quark (gluon) momentum from incoming $\vec{p}_{1}$ ($\vec{p}_{2}$)
to outgoing $\vec{p}_{3}=\vec{p}_{1}-\vec{q}$ ($\vec{p}_{4}=\vec{p}_{2}+\vec{q}$\;),
\begin{equation}
	\begin{aligned}\label{eq:IndiTraRate}
		\omega(\vec{p}_{1},\vec{q},T)=&\int_{p_{2}} n_{B}(E_{2}) \bar{n}_{F}(E_{3}) \bar{n}_{B}(E_{4}) \cdot v_{rel}d\sigma.
	\end{aligned}
\end{equation}
We neglect the thermal effects on the outgoing particles and thus $\bar{n}_{F}(E_{3})\approx\bar{n}_{B}(E_{4})\approx 1$.
Summing over the polarization and color of the outgoing partons and averaging over those of incoming partons,
the differential production cross section for quark-gluon scattering can be expressed as
\begin{equation}
	\begin{aligned}\label{eq:DiffCroSec}
		v_{rel}d\sigma(\vec{p}_{1},\vec{p}_{2}\to \vec{p}_{3},\vec{p}_{4}) &= \frac{1}{2E_{1}} \frac{1}{2E_{2}}
		\frac{d^{3}\vec{p}_{3}}{(2\pi)^{3}2E_{3}} \frac{d^{3}\vec{p}_{4}}{(2\pi)^{3}2E_{4}} \overline{|\mathcal{M}^{2}}|
		(2\pi)^{4} \delta^{(4)}(P_{in}-P_{out})
	\end{aligned}
\end{equation}
where, $v_{rel}=(\sqrt{(P_{1}P_{2})^{2}-(m_{1}m_{2})^{2}})/(E_{1}E_{2})$
is the relative velocity between the incoming HQ and thermal gluon.
The relevant scattering rate reads as
\begin{equation}
	\begin{aligned}\label{eq:IndiGammaHard}
		\Gamma^{hard}(E_{1},T) &= \int d^{3}\vec{q}~\omega(\vec{p}_{1},\vec{q}, T) \\ 
		&= \frac{1}{2E_{1}} \int_{p_{2}} \frac{n_{B}(E_{2})}{2E_{2}}
		\int_{p_{3}} \frac{1}{2E_{3}} \int_{p_{4}} \frac{1}{2E_{4}} \overline{|\mathcal{M}^{2}}|
		(2\pi)^{4} \delta^{(4)}(P_{in}-P_{out})
	\end{aligned}
\end{equation}

The energy loss and momentum diffusion coefficients can be obtained by inserting Eq.~(\ref{eq:IndiGammaHard})
into Eqs.~(\ref{eq:ELoss_Def}), (\ref{eq:KappaT_Def}) and (\ref{eq:KappaL_Def}), respectively.
Then, with the help of the $\delta$ function in Eq.~(\ref{eq:IndiGammaHard}),
we can reduce its integral down to three-dimensions in the numerical calculations, yielding~\cite{Peng:2024zvf, Li:2021nim}
\begin{equation}
	\begin{aligned}\label{eq:dEdz_Hard_t_Perturbative}
		\biggr[-\frac{dE}{dz}\biggr]^{hard}_{(t)} &= \frac{1}{256\pi^{3}p_{1}^{2}} \int_{p_{2,min}}^{\infty}dp_{2} E_{2} n_{B}(E_{2})
		\int_{-1}^{cos\psi|_{max}} d(cos\psi) \int_{t_{min}}^{t^{*}}dt \frac{b}{a^{3}} \; \overline{|\mathcal{M}^{2}|}_{(t)},
	\end{aligned}
\end{equation}
\begin{equation}
	\begin{aligned}\label{eq:KappaT_Hard_t_Perturbative}
		\biggr[\kappa_{T}\biggr]^{hard}_{(t)} =& \frac{1}{256\pi^{3}p_{1}^{3}E_{1}}
		\int_{p_{2,min}}^{\infty}dp_{2} E_{2} n_{B}(E_{2}) \int_{-1}^{cos\psi|_{max}} d(cos\psi) \\
		& \int_{t_{min}}^{t^{\ast}}dt \frac{1}{a} \biggr[ -\frac{m_{1}^{2}(D+2b^{2})}{8a^{4}} + \frac{E_{1}tb}{2a^{2}} - t(p_{1}^{2}+\frac{t}{4}) \biggr] \; \overline{|\mathcal{M}^{2}|}_{(t)},
	\end{aligned}
\end{equation}
and
\begin{equation}
	\begin{aligned}\label{eq:KappaL_Hard_t_Perturbative}
		\biggr[\kappa_{L}\biggr]^{hard}_{(t)} =& \frac{1}{256\pi^{3}p_{1}^{3}E_{1}}
		\int_{p_{2,min}}^{\infty}dp_{2} E_{2} n_{B}(E_{2}) \int_{-1}^{cos\psi|_{max}} d(cos\psi) \\
		& \int_{t_{min}}^{t^{\ast}} dt \frac{1}{a} \biggr[ \frac{E_{1}^{2}(D+2b^{2})}{4a^{4}}
		- \frac{E_{1}tb}{a^{2}} + \frac{t^{2}}{2} \biggr] \; \overline{|\mathcal{M}^{2}|}_{(t)}.
	\end{aligned}
\end{equation}
The integral boundaries and the short notations are summarized below:
\begin{subequations}
	\begin{align}
		&p_{2,min} = \frac{ |t^{\ast}|+\sqrt{t^{\ast 2} + 4m_{1}^{2} |t^{\ast}|} }{4(E_{1}+p_{1})} \label{eq:App_P2Min}, \\
		&cos\psi|_{max} = min \biggr\{ 1,~\frac{E_{1}}{p_{1}} - \frac{|t^{\ast}| + \sqrt{t^{\ast 2} + 4m_{1}^{2} |t^{\ast}|}}{4p_{1}p_{2}} \biggr\} \label{eq:App_PhiMax}, \\
		&t_{min} = -\frac{(s-m_{1}^{2})^{2}}{s} \label{eq:App_tMin}, \\
		&a = \frac{s-m_{1}^{2}}{p_{1}} \label{eq:App_aVal}, \\
		&b =-\frac{2t}{p_{1}^{2}} \bigr[ E_{1}(s-m_{1}^{2})-E_{2}(s+m_{1}^{2}) \bigr] \label{eq:App_bVal}, \\
		&c =-\frac{t}{p_{1}^{2}} \biggr\{ t\bigr[ (E_{1}+E_{2})^{2}-s \bigr] + 4p_{1}^{2}p_{2}^{2}sin^{2}\psi \biggr\} \label{eq:App_cVal}, \\
		&D = b^{2}+4a^{2}c=-t \biggr[ ts + (s-m_{1}^{2})^{2} \biggr] \cdot \biggr( \frac{4E_{2}sin\psi}{p_{1}} \biggr)^{2} \label{eq:App_DVal}.
	\end{align}
\end{subequations}

The contributions from $s$- and $u$ channels (panels b and c in Fig.~\ref{fig:TwoBody_Diag})
are not divergent for small momentum transfers due to
the finite heavy quark mass ($m_{c}=1.5\rm~GeV$ and $m_{b}=4.75\rm~GeV$).
Therefore, the $su$ channel contributions for the energy loss and
momentum diffusion coefficients can be obtained directly by modifying
the integral boundaries in the $t$ channel [Eqs.~(\ref{eq:dEdz_Hard_t_Perturbative}),
(\ref{eq:KappaT_Hard_t_Perturbative}) and (\ref{eq:KappaL_Hard_t_Perturbative})]:
(1) the $|\vec{p}_{2}|$ integration down to zero, $|\vec{p}_{2}|_{min}=0$ in Eq.~(\ref{eq:App_P2Min});
(2) the $cos\psi$ integration up to unity, $cos\psi|_{max}=1$ in Eq.~(\ref{eq:App_PhiMax});
(3) the $t$ integration up to zero, $t^{\ast}=0$ in Eq.~(\ref{eq:App_tMin}).
We arrive at
\begin{equation}
	\begin{aligned}\label{eq:dEdz_Hard_su_Perturbative}
		\biggr[-\frac{dE}{dz}\biggr]^{hard}_{(su)} &= \frac{1}{256\pi^{3}p_{1}^{2}} \int_{0}^{\infty}dp_{2} E_{2} n_{B}(E_{2})
		\int_{-1}^{1} d(cos\psi) \int_{t_{min}}^{0}dt \frac{b}{a^{3}} \; \overline{|\mathcal{M}^{2}|}_{(su)},
	\end{aligned}
\end{equation}
\begin{equation}
	\begin{aligned}\label{eq:KappaT_Hard_su_Perturbative}
		\biggr[\kappa_{T}\biggr]^{hard}_{(su)} =& \frac{1}{256\pi^{3}p_{1}^{3}E_{1}}
		\int_{0}^{\infty}dp_{2} E_{2} n_{B}(E_{2}) \int_{-1}^{1} d(cos\psi) \\
		& \int_{t_{min}}^{0}dt \frac{1}{a} \biggr[ -\frac{m_{1}^{2}(D+2b^{2})}{8a^{4}} + \frac{E_{1}tb}{2a^{2}} - t(p_{1}^{2}+\frac{t}{4}) \biggr] \; \overline{|\mathcal{M}^{2}|}_{(su)},
	\end{aligned}
\end{equation}
\begin{equation}
	\begin{aligned}\label{eq:KappaL_Hard_su_Perturbative}
		\biggr[\kappa_{L}\biggr]^{hard}_{(su)} =& \frac{1}{256\pi^{3}p_{1}^{3}E_{1}}
		\int_{0}^{\infty}dp_{2} E_{2} n_{B}(E_{2}) \int_{-1}^{1} d(cos\psi) \\
		& \int_{t_{min}}^{0} dt \frac{1}{a} \biggr[ \frac{E_{1}^{2}(D+2b^{2})}{4a^{4}}
		- \frac{E_{1}tb}{a^{2}} + \frac{t^{2}}{2} \biggr] \; \overline{|\mathcal{M}^{2}|}_{(su)}.
	\end{aligned}
\end{equation}
The matrix elements in vacuum, $\overline{|\mathcal{M}^{2}|}$ for $t$-, $s$- and $u$ channels,
can be found in Ref.~\cite{Peng:2024zvf, Li:2021nim}.

\subsection{Background field effective theory in a semi-QGP}\label{subsec:EffectiveeModel}
This Subsection is dedicated to the calculation of the collisional energy loss and momentum diffusion coefficients of heavy quarks in a semi-QGP.
We work in a Polyakov loop background as introduced in Ref.~\cite{Hidaka:2009hs}.
The nonperturbative influence is encoded in the $T$ dependency of the Polyakov loop, in particular in the semi-QGP regime near the critical temperature $T_{c}<T\lesssim2-3T_{c}$, where the confinement is not complete, but the thermal gluons are not fully liberated.

To quantify the effect of the nontrivial Polyakov loop in the deconfining phase transition for $SU(N_{c})$ gauge theory, we take the classical background field as a constant, diagonal matrix for the temporal component of the vector potential~\cite{Hidaka:2009hs}
\begin{equation}
	\begin{aligned} \label{eq:ClasBKG}
		(A_{0}^{cl})_{ab}=\frac{1}{g}\mathcal{Q}^{a} \delta_{ab}, \qquad \mathcal{Q}^{a}\equiv 2\pi T Q^{a},
	\end{aligned}
\end{equation}
where $g$ is the coupling constant for the gauge field.
There is no background field for its spatial components $A_{i}$.
As an $SU(N_{c})$ matrix, the sum of $\mathcal{Q}^{a}$'s vanishes, $\sum_{a=1}^{N_{c}}\mathcal{Q}^{a}=0$.
For three colors ($N_{c}=3$), the real $\mathcal{Q}^{a}$ has only one independent variable $Q$,
\begin{equation}
	\begin{aligned} \label{eq:Qvector_Def}
		Q^{a} = (Q,0,-Q),
	\end{aligned}
\end{equation}
which is dimensionless.
The analytical expression for $Q(T)$ is obtained from its equation of motion, which is rigorously derived by minimizing the next-to-leading-order effective potential of the semi-QGP, as detailed in Ref.~\cite{Du:2024riq}.
It yields
\begin{equation}
	\begin{aligned} \label{eq:Qvector_vsT}
		Q(T) = \frac{1}{36}\biggr[9-\sqrt{81-80\biggr(\frac{T_{c}}{T}\biggr)^{2}}\biggr],
	\end{aligned}
\end{equation}
with the critical temperature $T_{c}=0.27~{\rm GeV}$.

The Wilson line in the temporal direction reads
\begin{equation}
	\begin{aligned} \label{eq:Gamma_BFET_WilsonLine}
		\mathbf{L} = \mathcal{P} {\rm exp} \biggr( ig\int_{0}^{\beta}d\tau A_{0}^{cl} \biggr),
	\end{aligned}
\end{equation}
where $\mathcal{P}$ is the Euclidean time ordering and $\beta\equiv 1/T$ is the inverse temperature.
The trace of the Wilson line is the gauge invariant first Polyakov loop,
\begin{equation}
	\begin{aligned} \label{eq:PolyaLoop_1}
		\ell = \frac{1}{N_{c}} Tr\mathbf{L} = \frac{1}{N_{c}} \sum_{a=1}^{N_{c}} {\rm exp}(i2\pi Q^{a}).
	\end{aligned}
\end{equation}    
For three colors ($N_{c}=3$), the summation over $a$ can be done exactly, giving
\begin{equation}
	\begin{aligned} \label{eq:PolyaLoop_2}
		\ell = \frac{1}{3} \biggr[ 1+2{\rm cos} \bigr( 2\pi Q\bigr) \biggr].
	\end{aligned}
\end{equation}
Note that in the confined phase ($T\approx T_{c}$), Eq.~(\ref{eq:Qvector_vsT}) gives $Q\approx 1/4$
and hence $\ell\approx 1/3$ from Eq.~(\ref{eq:PolyaLoop_2}) in the pure gauge theory.
In the deconfined phase ($T\gg T_{c}$), we have $Q\approx 0$ and $\ell\approx 1$.

In the double line basis~\cite{Hidaka:2009hs},
quarks in the fundamental representation have one color line,
so their color indices are denoted by a single symbol,
e.g. $\mathcal{Q}^{a}$ with color index $a\in\{1,2,\dots,N_{c}$\},
while gluons in the adjoint representation have two color lines,
so their color indicies are denoted by a pair fundamental indices,
e.g. $\mathcal{Q}^{ab}\equiv \mathcal{Q}^{a}-\mathcal{Q}^{b}$ with $a,b\in\{1,2,\dots,N_{c}\}$.
In this basis the Euclidean energy of a particle $p_{0}$
can be shifted by the background field~\cite{Hidaka:2008dr, Hidaka:2009hs}:
\begin{equation}
	\begin{aligned} \label{eq:QuarkEng_BFET}
		(P^{\mu})^{a} = (p_{0}^{a},\vec{p}\;) = (p_{0}+\mathcal{Q}^{a},\vec{p}\;)
	\end{aligned}
\end{equation}
for quarks and
\begin{equation}
	\begin{aligned} \label{eq:GluonEng_BFET}
		(P^{\mu})^{ab} = (p_{0}^{ab},\vec{p}\;) = (p_{0}+\mathcal{Q}^{ab},\vec{p}\;)
	\end{aligned}
\end{equation}
for gluons.
In the imaginary time formalism of the thermal field theory,
because of the usual boundary conditions in imaginary time,
the Euclidean energy is typically written as $p_{0}=i\omega_{n}$,
where $\omega_{n}$ is the Matsubara frequency:
$\omega_{n}=2n\pi T$ ($n\in \mathbb{Z}$)
for gluons in thermal equilibrium at a temperature $T$,
and $\omega_{n}=(2n+1)\pi T$ for quarks.
We note that, although the energies for quarks and
gluons are rather different in Euclidean spacetime,
the proper procedure for analytic continuation to Minkowski spacetime
is to continue the entire Euclidean energy to $-iE$,
where $E$ is a continuous energy variable~\cite{Hidaka:2009hs, Hidaka:2015ima}.

Using the double line notations, the color projector is
given by a product of Kronecker deltas,
\begin{equation}
	\begin{aligned} \label{eq:Projector_DoubleLine}
		\mathcal{P}_{cd}^{ab} = \delta_{c}^{a}\delta_{d}^{b}-\frac{1}{N_{c}}\delta^{ab}\delta_{cd},
	\end{aligned}
\end{equation}
which satisfies $\mathcal{P}_{cd}^{ab}=\mathcal{P}^{ab,dc}=\mathcal{P}_{ba,cd}$
and $\mathcal{P}_{ef}^{ab}\mathcal{P}_{cd}^{ef}=\mathcal{P}_{cd}^{ab}$.
The first term on the right-hand side of Eq.~(\ref{eq:Projector_DoubleLine})
shows the orthogonality between the $N_{c}^{2}-N_{c}$ off-diagonal
and $N_{c}-1$ diagonal gluons,
and the second term mixes $N_{c}-1$ different diagonal gluons,
which is due to the over completeness of the double line basis.
The relevant generators are expressed as $t^{ab}_{cd}=\frac{1}{\sqrt{2}}\mathcal{P}_{cd}^{ab}$,
normalized as $Tr(t^{ab}t^{cd})=\frac{1}{2}\mathcal{P}^{ab,cd}$.
See Ref.~\cite{Hidaka:2009hs} for more details about the double line basis.
The Feynman rules in double line basis can be found in Refs.~\cite{Hidaka:2009hs, Guo:2018scp}.

As discussed in Ref.~\cite{Meisinger:2001cq},
by taking into account the classical background field
and considering further a mass scale in the dispersion relation for gauge bosons,
$\omega_{q}=\sqrt{\vec{q}^{\;2}+M^{2}_{g}}$,
one can expand the resulting effective potential up to the next-to-leading order
in the high temperature limit, $M_{g}\ll T$.
It yields~\cite{Meisinger:2001cq, Hidaka:2020vna}
\begin{equation}
	\begin{aligned}\label{eq:EffectPotential_BFET}
		\mathcal{V}_{eff} &= \mathcal{V}_{pt} + \mathcal{V}_{npt} = \frac{2\pi^{2}T^{4}}{3}\sum_{ab}\mathcal{P}^{ab,ba}B_{4}(\mid Q^{ab}\mid) + \frac{M^{2}_{g}T^{2}}{2}\sum_{ab}\mathcal{P}^{ab,ba}B_{2}(\mid Q^{ab}\mid),
	\end{aligned}
\end{equation}
with the second and fourth periodic Bernoulli polynomials given by
$B_{2}(x)=x^{2}-x+1/6$ and $B_{4}(x)=x^{2}(1-x)^{2}-1/30$~\cite{Guo:2018scp}, respectively;
the mass scale is fixed via $M_{g}/T_{c}=2\sqrt{10}\pi/9$~\cite{Du:2024riq}.
The first term on the right-hand side of Eq.~(\ref{eq:EffectPotential_BFET}), $\mathcal{V}_{pt}$,
denotes the result obtained at leading order.
The corresponding equation of motion of the background field shows that the thermal system
is always in a fully deconfined phase with vanishing background field $\mathcal{Q}=0$,
and no phase transition can happen~\cite{Du:2024riq}.
The second one, $\mathcal{V}_{npt}$,
represents the next-to-leading order term,
which drives the system toward confinement,
particularly near the critical temperature regime.

It is argued~\cite{Hidaka:2008dr, Hidaka:2009hs} that,
despite working with a background field,
the energy-momentum relation of a single, on-shell particle in
Minkowski spacetime remains unchanged $E=\sqrt{\vec{p}^{\;2}+m^{2}}$.
It means that, for a given scattering process, the structure of its matrix elements
and the associated kinematic integrals is not altered fundamentally by the background field $\mathcal{Q}$,
only the statistical occupancy is.
For processes in which all the particle's momenta are hard $|\vec{p}\;|\sim T$,
the loop corrections such as the thermal masses or resummed propagators are
subleading and thus neglected.
So, the only significant impact of the background field is through the
modification of the particle's thermal distribution functions,
which now include the $\mathcal{Q}$-dependent chemical-like phases.
For processes in which some momenta are soft $|\vec{p}\;|\sim gT$ ($g\ll 1$),
where the medium-induced loop corrections become important,
the background field affects both the HTL self-energies and effective propagators.
So, it is also necessary to include the $\mathcal{Q}$ dependency
of the hard thermal loops.
We will introduce these aspects in details in the following.
The perturbative calculations (see Sec.~\ref{subsec:PerturbativeModel})
will be extended to include the nontrivial nonperturbative contributions
in the semi-QGP regime.

\subsubsection{Thermal distribution functions and effective propagators in a background field}\label{susubbsec:BFET_ThermalEffProp}
In this part we introduce some useful tricks for computing scattering rate.
As discussed in Ref.~\cite{Hidaka:2009hs},
by expanding the propagators in a mixed representation~\cite{Kapusta:2006pm, Bellac:2011kqa}
in the presence of a background field $\mathcal{Q}\ne 0$,
and then comparing with the ones for $\mathcal{Q}=0$,
it is realized that the only change for $\mathcal{Q}\ne 0$ is the modification
in the thermal distribution functions,
which are now expressed as,
\begin{equation}
	\begin{aligned} \label{eq:BFET_ThermalDis_Boson}
		n_{B}(E\pm i\mathcal{Q}^{ab}) = \frac{1}{e^{\beta \bigr(E \pm i\mathcal{Q}^{ab}\bigr)} - 1},\\
	\end{aligned}
\end{equation}
for bosons, and
\begin{equation}
	\begin{aligned} \label{eq:BFET_ThermalDis_Fermi}
		n_{F}(E\pm i\mathcal{Q}^{a}) = \frac{1}{e^{\beta \bigr(E \pm i\mathcal{Q}^{a}\bigr)} + 1},\\
	\end{aligned}
\end{equation}
for fermions.
When $\mathcal{Q}=0$, Eqs.~(\ref{eq:BFET_ThermalDis_Boson}) and (\ref{eq:BFET_ThermalDis_Fermi})
are reduced to Eqs.~(\ref{eq:ThermalDis_Boson}) and (\ref{eq:ThermalDis_Fermi}), respectively.
Consequently, the background field induces an energy shift in the gluon thermal distribution,
which is operationally equivalent to introducing an imaginary chemical potential $i\mathcal{Q}$ for color.
Similar to Eq.~(\ref{eq:ThermalDis_StatEff}), we define
\begin{subequations}
	\begin{align}
		\bar{n}_{B}(E\pm i\mathcal{Q}^{ab}) =& 1 + n_{B}(E\pm i\mathcal{Q}^{ab}) \label{eq:BFET_ThermalDis_StatEff_Boson}, \\
		\bar{n}_{F}(E\pm i\mathcal{Q}^{a}) =& 1 - n_{F}(E\pm i\mathcal{Q}^{a}) \label{eq:BFET_ThermalDis_StatEff_Fermi},
	\end{align}
\end{subequations}
which accounts for the statistical effects for the outgoing particles.

The color-averaged gluon distribution becomes~\cite{Singh:2018wps}
\begin{equation}
	\begin{aligned} \label{eq:BFET_ThermalDis_Boson_Avg}
		n_{Avg,B}(E,T) &= \frac{1}{N_{c}^2} \sum_{a,b=1}^{N_{c}} n_{B}(E- i\mathcal{Q}^{ab})
		= \frac{1}{N_{c}^2} \sum_{a,b=1}^{N_{c}} n_{B}(E+ i\mathcal{Q}^{ab}) \\
		&= \frac{1}{9} \biggr[ \frac{3}{e^{\beta E}-1} + \frac{e^{\beta E}(6\ell-2)-4}{1+e^{2\beta E}+e^{\beta E}(1-3\ell)} + \frac{e^{\beta E}(9\ell^{2}-6\ell-1)-2}{1+e^{2\beta E}+e^{\beta E}(1+6\ell-9\ell^{2})} \biggr]
	\end{aligned}
\end{equation}
for three colors ($N_{c}=3$).
The influence of the background field on the distribution functions,
can be quantified by comparing the color-averaged distributions
in the confined ($\ell=0$) and deconfined phases ($\ell=1$),
\begin{equation}
	\begin{aligned} \label{eq:BFET_ThermalDis_Ratio}
		\frac{n_{Avg,B}|_{\ell=0}}{n_{Avg,B}|_{\ell=1}} &= \frac{e^{\beta E}-1}{e^{3\beta E}-1}<1.
	\end{aligned}
\end{equation}
The denominator $n_{Avg,B}|_{\ell=1}=1/(e^{\beta E}-1)$ is nothing but
the thermal distribuion for $\mathcal{Q}=0$, as shown in Eq.~(\ref{eq:ThermalDis_Boson}).
See Fig.~\ref{fig:Ratio_nB} for further discussions.

The HTL-resummed gluon propagator in the effective theory, $D_{\mu\nu}^{de,fg}$,
was investigated in Ref.~\cite{Guo:2020jvc},
and the obtained results were further utilized to study the influence of the nontrivial Polyakov loop
on the in-medium properties of the heavy quark~\cite{Du:2024riq}
and heavy quarkonium states~\cite{Liu:2024fki} in a semi-QGP.
We employ this resummed propagator in this work.
In the double line basis the $N^{2}_{c}-N_{c}$ off-diagonal components are expressed as
\begin{subequations}
	\begin{align}
		\Delta^{de,fg}_{T,off-diag}(q_{0},\vec{q}\;) &\stackrel{d\ne e}{=} \delta^{dg}\delta^{ef} \frac{-1}{q_{0}^{2}-\vec{q}^{\;2}-(\mathcal{M}_{D}^{2})^{de}_{off-diag}\Pi_{T}} \label{eq:BFET_GluonPropagator_OffDiag_T}, \\
		\Delta^{de,fg}_{L,off-diag}(q_{0},\vec{q}\;) &\stackrel{d\ne e}{=} \delta^{dg}\delta^{ef}\frac{-1}{\vec{q}^{\;2}+(\mathcal{M}_{D}^{2})^{de}_{off-diag}\Pi_{L}} \label{eq:BFET_GluonPropagator_OffDiag_L},
	\end{align}
\end{subequations}
where, $\delta^{dg}$ and $\delta^{ef}$ are Kronecker deltas related to the color indices.
$(\mathcal{M}_{D}^{2})^{de}_{off-diag}$ is the $\mathcal{Q}$-modified Debye screening mass squared
\begin{equation}
	\begin{aligned} \label{eq:BFET_OffDiag_DebMas}
		(\mathcal{M}_{D}^{2})^{de}_{off-diag} &= \lambda^{de} m_{D}^{2} = \biggr\{ \frac{3}{N_{c}}\sum_{h=1}^{N_{c}}\left[B_{2}(|Q^{dh}|)+B_{2}(|Q^{he}|)\right]+\frac{10}{27}\left(\frac{T_{c}}{T}\right)^{2} \biggr\} m_{D}^{2},
	\end{aligned}
\end{equation}
with $m_{D}$ the Debye screening mass for vanishing background [Eq.~(\ref{eq:DebMas})].
The background field effect on the resummed propagators
is quantified by the parameter $\lambda^{de}(T)$ [Eq.~(\ref{eq:BFET_OffDiag_DebMas})],
which is a monotonically increasing function of temperature
that approaches unity at high temperature $\lambda^{de}(T\gg T_{c})=1$.
In this case, the obtained results for $\mathcal{Q}\ne 0$,
such as the effective propagators [Eqs.~(\ref{eq:BFET_GluonPropagator_OffDiag_T}) and (\ref{eq:BFET_GluonPropagator_OffDiag_L})],
and Debye screening mass [Eq.~(\ref{eq:BFET_OffDiag_DebMas})]
can reduce to the corresponding ones for $\mathcal{Q}= 0$ [Eqs.~(\ref{eq:Propagator_Soft_T}), (\ref{eq:Propagator_Soft_L}) and (\ref{eq:DebMas})].

The $N_{c}-1$ independent diagonal components of the
HTL-resummed gluon propagator are given by
\begin{subequations}
	\begin{align}
		\sum_{d,f=1}^{N_{c}-1} \mathcal{P}^{dd,ff} \Delta_{T,diag}^{dd,ff} &=
		\sum_{h=1}^{N_{c}-1} \frac{-1}{q_{0}^{2}-\vec{q}^{\;2}-(\mathcal{M}_{D}^{2})^{h}_{diag}\Pi_{T}} \label{eq:BFET_GluonPropagator_Diag_T}, \\
		\sum_{d,f=1}^{N_{c}-1} \mathcal{P}^{dd,ff} \Delta_{L,diag}^{dd,ff} &=
		\sum_{h=1}^{N_{c}-1} \frac{-1}{\vec{q}^{\;2}+(\mathcal{M}_{D}^{2})^{h}_{diag}\Pi_{L}} \label{eq:BFET_GluonPropagator_Diag_L},
	\end{align}
\end{subequations}
where $(\mathcal{M}_{D}^{2})^{h}_{diag}$ is also the $\mathcal{Q}$-modified Debye screening mass squared
\begin{equation}
	\begin{aligned} \label{eq:BFET_Diag_DebMas}
		(\mathcal{M}_{D}^{2})^{h}_{diag} &= \lambda^{h} m_{D}^{2} = \biggr[ 1+ \frac{10}{27}\Bigr(\frac{T_{c}}{T}\Bigr)^{2} + \frac{6}{N_{c}}\mathcal{F}^{h} \biggr] m_{D}^{2}.
	\end{aligned}
\end{equation}
The explicit form of the parameter $\mathcal{F}^{h}$ reads as~\cite{Guo:2020jvc}
\begin{equation}
	\begin{aligned} \label{eq:BFET_Diag_Alpha}
		\mathcal{F}^{1} = 3Q^{2}-3Q, \qquad \mathcal{F}^{2} = 9Q^{2}-5Q,
	\end{aligned}
\end{equation}
for $SU(N_{c}=3)$.
One can make similar discussions to those of the off-diagonal components.
The same conclusions can be drawn for the results based on the diagonal components.

We can see that $N_{c}^{2}-N_{c}$ off-diagonal gluons [Eqs.~(\ref{eq:BFET_GluonPropagator_OffDiag_T}) and (\ref{eq:BFET_GluonPropagator_OffDiag_L})]
and $N_{c}-1$ diagonal ones [Eqs.~(\ref{eq:BFET_GluonPropagator_Diag_T}) and (\ref{eq:BFET_GluonPropagator_Diag_L})]
acquire different $\mathcal{Q}$ dependent modifications,
and thus become distinguishable by their associated screening mass [Eqs.~(\ref{eq:BFET_OffDiag_DebMas}) and (\ref{eq:BFET_Diag_DebMas})].
To obtain the heavy quark energy loss and momentum diffusion coefficients in the presence of a background field,
we can take as reference the perturbative strategy as introduced in Sec.~\ref{subsec:PerturbativeModel},
and then extend it by: (1) replacing the thermal distribution function for
$\mathcal{Q}=0$ [Eqs.~(\ref{eq:ThermalDis_Boson}), (\ref{eq:ThermalDis_Fermi}) and (\ref{eq:ThermalDis_StatEff})]
with the $\mathcal{Q}$-depedent ones [Eqs.~(\ref{eq:BFET_ThermalDis_Boson}), (\ref{eq:BFET_ThermalDis_Fermi}), (\ref{eq:BFET_ThermalDis_StatEff_Boson}) and (\ref{eq:BFET_ThermalDis_StatEff_Boson})];
(2) replacing the HTL-resummed gluon propagator for $\mathcal{Q}=0$ [Eqs.~(\ref{eq:Propagator_Soft_T}) and (\ref{eq:Propagator_Soft_L})] with
the ones for $\mathcal{Q}\ne 0$ [Eqs.~(\ref{eq:BFET_GluonPropagator_OffDiag_T}), (\ref{eq:BFET_GluonPropagator_OffDiag_L}), (\ref{eq:BFET_GluonPropagator_Diag_T}) and (\ref{eq:BFET_GluonPropagator_Diag_L})].

\subsubsection{The results in soft region}\label{susubbsec:BFET_Soft}
In soft collisions the color-averaged scattering rate of the heavy quark becomes
\begin{equation}
	\begin{aligned}\label{eq:Gamma_ColorAvg_BFET}
		\Gamma^{\mathcal{Q}\ne0;soft}_{(t)}(E_{1},T) = \frac{1}{N_{c}}\sum_{a=1}^{N_{c}}\bigr[\Gamma^{\mathcal{Q}\ne0;soft}_{(t)}\bigr]^{a}(E_{1},T),
	\end{aligned}
\end{equation}
where, $N_{c}$ is the color factor of the incoming heavy quark.
By extending Weldon’s model~\cite{WeldonPRD83, PhysRevD.44.1298} close to the critical temperature,
the result with a given color $a$ can be obtained from Eq.~(\ref{eq:Gamma_Soft1}),
\begin{equation}
	\begin{aligned} \label{eq:Gamma_BFET_Soft}
		\bigr[\Gamma^{\mathcal{Q}\ne0;soft}_{(t)}\bigr]^{a}(E_{1},T) &= -\frac{1}{2E_{1}} \bar{n}_{F}(E_{1}-i\mathcal{Q}^{a})
		Tr \bigr[ (\slashed{P}^{a}_{1} + m_{1})\cdot{Im}\Sigma_{ab}^{de,fg} (E_{1}+i\epsilon,\vec{p}_{1}) \bigr],
	\end{aligned}
\end{equation}
with the four-momentum $P^{a}_{1}=(p_{1;0}^{a},\vec{p}\;)=(p_{1;0}+i\mathcal{Q}^{a},\vec{p}\;)$ of the incoming heavy quark.

Following a similar strategy as described in Sec.~\ref{susubbsec:Pert_Soft},
we can calculate the heavy quark self-energy in terms of the mixed representations of the bare quark and HTL-resummed gluon propagators.
Detailed derivations can be found in Appendix~\ref{appendix:Gamma_Soft_BFET}.
The resulting interaction rate is contributed by the off-diagonal[Eq.~(\ref{eq:Gamma_ColorAvg_BFET_Soft_OffDiag})]
and diagonal gluons [Eq.~(\ref{eq:Gamma_ColorAvg_BFET_Soft_Diag})],
\begin{equation}
	\begin{aligned}\label{eq:Gamma_ColorAvg_BFET_Soft}
		\Gamma^{\mathcal{Q}\ne0;soft}_{(t)}(E_{1},T) &= \bigr[\Gamma^{\mathcal{Q}\ne0;soft}_{(t)}\bigr]_{off-diag} + \bigr[\Gamma^{\mathcal{Q}\ne0;soft}_{(t)}\bigr]_{diag} \\
		& = \frac{g^{2}}{2N_{c}} \int_q \int d\omega \;
		\delta(\omega-\vec{v}_{1}\cdot\vec{q}\;)
		\biggr\{ \mathcal{B}_{L} + \vec{v}_{1}^{\;2}\bigr[ 1-(\widehat{v}_{1}\cdot\widehat{q})^{2} \bigr] \mathcal{B}_{T} \biggr\},
	\end{aligned}
\end{equation}
with
\begin{equation}
	\begin{aligned} \label{eq:BFET_EffRhoTL1}
		\mathcal{B}_{T/L}(\omega,q) &= \sum_{\substack{d,e=1 \\ (d\ne e)}}^{N_{c}} \mathcal{N}(\omega,\mathcal{Q}^{de}) \cdot \rho_{T/L,off-diag}^{de,ed} + \sum_{h=1}^{N_{c}-1}\bar{n}_{B}(\omega) \cdot \rho^{h}_{T/L,diag}.
	\end{aligned}
\end{equation}
The spectral functions for the off-diagonal, $\rho_{T/L,off-diag}^{de,ed}$, and diagonal gluons, $\rho^{h}_{T/L,diag}$,
are shown in Eqs.~(\ref{eq:BFET_GammaRhoT_Soft_OffDiag}), (\ref{eq:BFET_GammaRhoL_Soft_OffDiag}),
(\ref{eq:BFET_GammaRhoT_Soft_Diag}) and (\ref{eq:BFET_GammaRhoL_Soft_Diag}).
The factor $\mathcal{N}(\omega,\mathcal{Q}^{de})$ indicates
a background-field correction [Eq.~(\ref{eq:Imag_Approx2})].
By comparing the scattering rate for $\mathcal{Q}\ne0$ [Eq.~(\ref{eq:Gamma_ColorAvg_BFET_Soft})]
with that for $\mathcal{Q}=0$ [Eq.~(\ref{eq:Trace_Soft7})],
it is found that,
the modification due to the background field is simply given by
replacing the relevant variables for $\mathcal{Q}=0$:
the color factors $C_{F}\to 1/(2N_{c})$,
the weighted spectral distributions $\mathcal{A}_{T/L}\to\mathcal{B}_{T/L}$
and the Debye screening masses $m_{D}\to \mathcal{M}_{D}$.

With Eq.~(\ref{eq:Gamma_ColorAvg_BFET_Soft})
we can carry out the polar angular integral over $\vec{q}$,
and the resulting energy loss [Eq.~(\ref{eq:ELoss_Def})] and
momentun diffusions [Eqs.~(\ref{eq:KappaT_Def}) and (\ref{eq:KappaL_Def})] take the forms
\begin{equation}
	\begin{aligned}\label{eq:dEdz_Soft_BFET}
		\biggr[-\frac{dE}{dz}\biggr]^{\mathcal{Q}\ne0;\;soft}_{(t)} =& \frac{g^{2}}{16\pi^{2}N_{c}v^{2}_{1}}
		\int^{0}_{t^{*}} dt \; (-t) \int_{0}^{v_{1}}dx \frac{x}{(1-x^{2})^{2}}
		\bigr[ \widetilde{\mathcal{B}}_{L-} + (v_{1}^{2}-x^{2})\widetilde{\mathcal{B}}_{T-} \bigr],
	\end{aligned}
\end{equation}
\begin{equation}
	\begin{aligned}\label{eq:KappaT_Soft_BFET}
		\biggr[\kappa_{T}\biggr]^{\mathcal{Q}\ne0;\;soft}_{(t)} =& \frac{g^{2}}{32\pi^{2}N_{c}v_{1}^{3}} \int^{0}_{t^{\ast}} dt \; (-t)^{3/2}
		\int_{0}^{v_{1}}dx \frac{v_{1}^{2}-x^{2}}{(1-x^{2})^{5/2}}
		\bigr[ \widetilde{\mathcal{B}}_{L+} + (v_{1}^{2}-x^{2})\widetilde{\mathcal{B}}_{T+} \bigr],
	\end{aligned}
\end{equation}
and
\begin{equation}
	\begin{aligned}\label{eq:KappaL_Soft_BFET}
		\biggr[\kappa_{L}\biggr]^{\mathcal{Q}\ne0;\;soft}_{(t)} =& \frac{g^{2}}{16\pi^{2}N_{c}v_{1}^{3}} \int^{0}_{t^{\ast}} dt \; (-t)^{3/2}
		\int_{0}^{v_{1}}dx \frac{x^{2}}{(1-x^{2})^{5/2}}
		\bigr[ \widetilde{\mathcal{B}}_{L+} + (v_{1}^{2}-x^{2})\widetilde{\mathcal{B}}_{T+} \bigr],
	\end{aligned}
\end{equation}
with $t=\omega^2-q^{2}$ and $x=\omega/q$.
The factor $\widetilde{\mathcal{B}}_{T/L\pm}$ reads
\begin{equation}
	\begin{aligned} \label{eq:Pert_SH2}
		\widetilde{\mathcal{B}}_{T/L\pm} &= \mathcal{B}_{T/L}(\omega) \pm \mathcal{B}_{T/L}(-\omega),
	\end{aligned}
\end{equation}
which can be further rewritten as
\begin{subequations}
	\begin{align}
		\widetilde{\mathcal{B}}_{T/L-} &= \sum_{\substack{d,e=1 \\ (d\ne e)}}^{N_{c}} \rho_{T/L,off-diag}^{de,ed} + \sum_{h=1}^{N_{c}-1}\rho^{h}_{T/L,diag}, \label{eq:BFET_Bmfactor2} \\ 
		\widetilde{\mathcal{B}}_{T/L+} &= coth\left(\frac{\omega}{2T}\right) \cdot
		\biggr\{ \sum_{\substack{d,e=1 \\ (d\ne e)}}^{N_{c}} \rho_{T/L,off-diag}^{de,ed} \biggr/
		\biggr[ 1+\frac{1-cos(\beta\mathcal{Q}^{de})}{2sinh^{2}(\frac{\omega}{2T})} \biggr]
		+ \sum_{h=1}^{N_{c}-1}\rho^{h}_{T/L,diag} \biggr\}, \label{eq:BFET_Bpfactor2}
	\end{align}
\end{subequations}

Note that the spectral functions $\rho_{T/L}(\omega)$ are odd functions of $\omega$.
The identities,
\begin{subequations}
	\begin{align}
		\mathcal{N}(\omega,\mathcal{Q}^{de}) + \mathcal{N}(-\omega,\mathcal{Q}^{de}) &= 1, \label{eq:Ident_NPlus} \\
		\bar{n}_{B}(\omega) + \bar{n}_{B}(-\omega) &= 1, \label{eq:Ident_nBbarPlus}
	\end{align}
\end{subequations}
and
\begin{subequations}
	\begin{align}
		\mathcal{N}(\omega,\mathcal{Q}^{de}) - \mathcal{N}(-\omega,\mathcal{Q}^{de}) &= coth\left(\frac{\omega}{2T}\right) \biggr/ \biggr[ 1+\frac{1-cos(\beta\mathcal{Q}^{de})}{2sinh^{2}(\frac{\omega}{2T})} \biggr], \label{eq:Ident_NMinus} \\
		\bar{n}_{B}(\omega) - \bar{n}_{B}(-\omega) &= coth\left(\frac{\omega}{2T}\right), \label{eq:Ident_nBbarMinus}
	\end{align}
\end{subequations}
are used to obtain Eqs.~(\ref{eq:BFET_Bmfactor2}) and (\ref{eq:BFET_Bpfactor2}), respectively.

\subsubsection{The results in hard region}\label{susubbsec:BFET_Hard}
In the hard momentum region, where all relevant momenta are of order temperature, $|\vec{p}\;|\gtrsim T\gg gT$,
the effect of the background temporal gluon field $\mathcal{Q} \sim T$,
at leading order in the QCD coupling $g$,
manifests primarily through modifications of the statistical distribution functions.
In this case $\mathcal{Q}$ acts as an imaginary color-dependent chemical potential,
its presence alters the thermal weights of both gluons [Eq.~(\ref{eq:BFET_ThermalDis_Boson})]
and quarks [Eq.~(\ref{eq:BFET_ThermalDis_Fermi})].
However, the $\mathcal{Q}$-dependent corrections to the kinematic structure of scattering matrix elements
appear only at higher orders in the coupling, typically $\mathcal{O}(g^2)$,
which lie beyond the scope of a leading order analysis~\cite{Hidaka:2015ima, JeanPR02, Hidaka:2008dr}.
Consequently, for tree-level $2 \to 2$ processes involving hard external momenta,
it is consistent to retain the standard matrix elements computed in the trivial background and
incorporate the background field solely through the modified thermal distribution functions,
as well as the $\mathcal{Q}$-dependent color factors.

In this part we will employ such $\mathcal{Q}$-dependent thermal distributions and color factors
to calculate the heavy quark scattering rate, energy loss
and momentum diffusion coefficients in hard collisions at leading order.
Considering a given channel of heavy quark scattering off gluons for $\mathcal{Q}\ne0$,
the relevant kinematic structure of the scattering amplitude is identical to that for $\mathcal{Q}=0$,
while the color structure will be modified by a $\mathcal{Q}$-dependent factor.

We now can calculate the heavy quark scattering rate and
the derived energy loss and momentum diffusion coefficients in hard collisions.
The general strategy contains the thermal distributions ($n$ for incoming and $\bar{n}$ for outgoing particles),
which usually prevent the calculations of the relevant integrals.
For simplicity, we neglect the thermal effects on the final states by replacing $\bar{n}\to 1$.
The color-averaged interaction rate of heavy quark scattering off gluons can be obtained from Eq.~(\ref{eq:IndiGammaHard})
by using the $\mathcal{Q}$-dependent thermal distribution functions,
\begin{equation}
	\begin{aligned} \label{eq:Gamma_Def_BFET_Hard1}
		\Gamma^{\mathcal{Q}\ne0;~hard}_{(t)}(E_{1},T) =& \frac{1}{2E_{1}} \ \biggr[ \frac{1}{8}\sum_{j,k=1}^{3}\left(1-\frac{1}{3}\delta^{jk}\right) \biggr]
		\int_{p_{2}} \frac{\bigr[n_{B}(E_{2}-i\mathcal{Q}^{jk})+n_{B}(E_{2}+i\mathcal{Q}^{jk})\bigr]\bigr/2}{2E_{2}} \\
		& \int_{p_{3}} \frac{1}{2E_{3}} \int_{p_{4}} \frac{1}{2E_{4}} \overline{|\mathcal{M}^{2}|}^{\mathcal{Q}=0}_{(t)}
		(2\pi)^{4} \delta^{(4)}(P_{in}-P_{out}).
	\end{aligned}
\end{equation}
The scattering amplitude for quark-gluon scattering
in the $t$ channel for $\mathcal{Q}\ne0$ is given by the term
$\frac{1}{8}\sum_{j,k=1}^{3}(1-\frac{1}{3}\delta^{jk}) \overline{|\mathcal{M}^{2}|}^{\mathcal{Q}=0}_{(t)}$,
where, $\overline{|\mathcal{M}^{2}|}^{\mathcal{Q}=0}_{(t)}$ is the relevant one
obtained in the absence of the background field $\mathcal{Q}=0$ [Eq.~(\ref{eq:MforQ_t})].
More detailed aspects are relegated to Appendix~\ref{appendix:ColorFactor_BFET}.
Similar results can be obtained in $s$- and $u$ channels by replacing the relevant scattering matrix for $\mathcal{Q}=0$.
Equqtion~(\ref{eq:Gamma_Def_BFET_Hard1}) can be reduced to the result in the absence of a
background field [Eq.~(\ref{eq:IndiGammaHard})] by setting $\mathcal{Q}=0$.

Inserting Eq.~(\ref{eq:Gamma_Def_BFET_Hard1}) into Eqs.~(\ref{eq:ELoss_Def}), (\ref{eq:KappaT_Def}) and (\ref{eq:KappaL_Def}), respectively,
we arrive at
\begin{equation}
	\begin{aligned} \label{eq:dEdz_BFET_Hard_t}
		\biggr[-\frac{dE}{dz}\biggr]^{\mathcal{Q}\ne0;~hard}_{(t)} =&
		\frac{1}{256\pi^{3}p_{1}^{2}} \biggr[\frac{1}{8}\sum_{j,k=1}^{3}\left(1-\frac{1}{3}\delta^{jk}\right) \biggr]
		\int_{p_{2,min}}^{\infty}dp_{2} E_{2} \frac{n_{B}(E_{2}-i\mathcal{Q}^{jk})+n_{B}(E_{2}+i\mathcal{Q}^{jk})}{2} \\
		& \int_{-1}^{cos\psi|_{max}} d(cos\psi) \int_{t_{min}}^{t^{*}}dt
		\frac{b}{a^{3}} \; \overline{|\mathcal{M}^{2}|}^{\mathcal{Q}=0}_{(t)},
	\end{aligned}
\end{equation}
\begin{equation}
	\begin{aligned} \label{eq:KappaT_BFET_Hard_t}
		\biggr[\kappa_{T}\biggr]^{\mathcal{Q}\ne0;~hard}_{(t)} =&
		\frac{1}{256\pi^{3}p_{1}^{3}E_{1}} \biggr[\frac{1}{8} \sum_{j,k=1}^{3}\left(1-\frac{1}{3}\delta^{jk}\right)\biggr]
		\int_{p_{2,min}}^{\infty}dp_{2} E_{2} 
		\frac{n_{B}(E_{2}-i\mathcal{Q}^{jk})+n_{B}(E_{2}+i\mathcal{Q}^{jk})}{2} \\
		&\int_{-1}^{cos\psi|_{max}} d(cos\psi) \int_{t_{min}}^{t^{\ast}}dt \frac{1}{a}
		\biggr[ -\frac{m_{1}^{2}(D+2b^{2})}{8a^{4}} + \frac{E_{1}tb}{2a^{2}} - t(p_{1}^{2}+\frac{t}{4}) \biggr]
		\cdot \overline{|\mathcal{M}^{2}|}^{\mathcal{Q}=0}_{(t)},
	\end{aligned}
\end{equation}
\begin{equation}
	\begin{aligned} \label{eq:KappaL_BFET_Hard_t}
		\biggr[\kappa_{L}\biggr]^{\mathcal{Q}\ne0;~hard}_{(t)} =&
		\frac{1}{256\pi^{3}p_{1}^{3}E_{1}} \biggr[\frac{1}{8} \sum_{j,k=1}^{3}\left(1-\frac{1}{3}\delta^{jk}\right)\biggr]
		\int_{p_{2,min}}^{\infty}dp_{2} E_{2} \frac{n_{B}(E_{2}-i\mathcal{Q}^{jk})+n_{B}(E_{2}+i\mathcal{Q}^{jk})}{2} \\
		& \int_{-1}^{cos\psi|_{max}} d(cos\psi) \int_{t_{min}}^{t^{\ast}} dt \frac{1}{a}
		\biggr[ \frac{E_{1}^{2}(D+2b^{2})}{4a^{4}} -\frac{E_{1}tb}{a^{2}} + \frac{t^{2}}{2} \biggr] \cdot \overline{|\mathcal{M}^{2}|}^{\mathcal{Q}=0}_{(t)},
	\end{aligned}
\end{equation}
with the integral boundaries and the short notations are shown in Eqs.~(\ref{eq:App_P2Min})-(\ref{eq:App_DVal}).
It appears that, comparing with the results for $\mathcal{Q}=0$
[Eqs.~(\ref{eq:dEdz_Hard_t_Perturbative}), (\ref{eq:KappaT_Hard_t_Perturbative}) and (\ref{eq:KappaL_Hard_t_Perturbative})],
the $\mathcal{Q}$-induced effects are entirely encapsulated by
modifications to the statistical distribution functions and the color factors.

Very similar to the situation for $\mathcal{Q}=0$
[Eqs.~(\ref{eq:dEdz_Hard_su_Perturbative}), (\ref{eq:KappaT_Hard_su_Perturbative}) and (\ref{eq:KappaL_Hard_su_Perturbative})],
the energy loss and the momentum diffusion coefficients from $s$- and $u$ channels (panels b and c in Fig.~\ref{fig:TwoBody_Diag_BFET})
can be obtained directly by modifying the integral boundaries of the relevant results in the $t$ channel,
\begin{equation}
	\begin{aligned} \label{eq:dEdz_BFET_Hard_su}
		\biggr[-\frac{dE}{dz}\biggr]^{\mathcal{Q}\ne0;~hard}_{(su)} =&
		\frac{1}{256\pi^{3}p_{1}^{2}} \biggr[\frac{1}{8}\sum_{j,k=1}^{3}\left(1-\frac{1}{3}\delta^{jk}\right) \biggr]
		\int_{0}^{\infty}dp_{2} E_{2} \frac{n_{B}(E_{2}-i\mathcal{Q}^{jk})+n_{B}(E_{2}+i\mathcal{Q}^{jk})}{2} \\
		& \int_{-1}^{1} d(cos\psi) \int_{t_{min}}^{0}dt
		\frac{b}{a^{3}} \; \overline{|\mathcal{M}^{2}|}^{\mathcal{Q}=0}_{(su)},
	\end{aligned}
\end{equation}
\begin{equation}
	\begin{aligned} \label{eq:KappaT_BFET_Hard_su}
		\biggr[\kappa_{T}\biggr]^{\mathcal{Q}\ne0;~hard}_{(su)} =&
		\frac{1}{256\pi^{3}p_{1}^{3}E_{1}} \biggr[\frac{1}{8} \sum_{j,k=1}^{3}\left(1-\frac{1}{3}\delta^{jk}\right)\biggr]
		\int_{0}^{\infty}dp_{2} E_{2} 
		\frac{n_{B}(E_{2}-i\mathcal{Q}^{jk})+n_{B}(E_{2}+i\mathcal{Q}^{jk})}{2} \\
		&\int_{-1}^{1} d(cos\psi) \int_{t_{min}}^{0}dt \frac{1}{a}
		\biggr[ -\frac{m_{1}^{2}(D+2b^{2})}{8a^{4}} + \frac{E_{1}tb}{2a^{2}} - t(p_{1}^{2}+\frac{t}{4}) \biggr]
		\cdot \overline{|\mathcal{M}^{2}|}^{\mathcal{Q}=0}_{(su)},
	\end{aligned}
\end{equation}
\begin{equation}
	\begin{aligned} \label{eq:KappaL_BFET_Hard_su}
		\biggr[\kappa_{L}\biggr]^{\mathcal{Q}\ne0;~hard}_{(su)} =&
		\frac{1}{256\pi^{3}p_{1}^{3}E_{1}} \biggr[\frac{1}{8} \sum_{j,k=1}^{3}\left(1-\frac{1}{3}\delta^{jk}\right)\biggr]
		\int_{0}^{\infty}dp_{2} E_{2} \frac{n_{B}(E_{2}-i\mathcal{Q}^{jk})+n_{B}(E_{2}+i\mathcal{Q}^{jk})}{2} \\
		& \int_{-1}^{1} d(cos\psi) \int_{t_{min}}^{0} dt \frac{1}{a}
		\biggr[ \frac{E_{1}^{2}(D+2b^{2})}{4a^{4}} -\frac{E_{1}tb}{a^{2}} + \frac{t^{2}}{2} \biggr] \cdot \overline{|\mathcal{M}^{2}|}^{\mathcal{Q}=0}_{(su)}.
	\end{aligned}
\end{equation}

Summing up the contributions from the soft collisions [Eqs.~(\ref{eq:dEdz_Soft_BFET}), (\ref{eq:KappaT_Soft_BFET}) and (\ref{eq:KappaL_Soft_BFET})],
hard collisions [Eqs.~(\ref{eq:dEdz_BFET_Hard_t}), (\ref{eq:KappaT_BFET_Hard_t}) and (\ref{eq:KappaL_BFET_Hard_t})],
and $su$ channels [Eqs.~(\ref{eq:dEdz_BFET_Hard_su}), (\ref{eq:KappaT_BFET_Hard_su}) and (\ref{eq:KappaL_BFET_Hard_su})]
we get the total energy loss and diffusion coefficients of a heavy quark scattering off gluons
in the background field.

\subsection{Implementation of the running coupling constant}\label{subsec:RunningCoupling}
To accurately capture the thermodynamic properties of the SU(3) gluon plasma across a wide temperature range, we adopt a temperature-dependent running coupling constant $\alpha_s(T)$ instead of a fixed value.
In our numerical calculations, we utilize the 1-loop running coupling formula given by:
\begin{equation}
        \begin{aligned}\label{eq:RunningCoupling}
	\alpha_s(\mu) = \frac{12\pi}{\left(11N_c - 2N_{f}\right) ln\left(\frac{\mu}{\Lambda_{QCD}}\right)^2},
        \end{aligned}
\end{equation}
where, $\Lambda_{QCD}=200~{\rm MeV}$ and $N_{f}$ is the number of active flavors in the plasma.
Consistent with our pure gauge framework, we set $N_f=0$ throughout this work.
The renormalization scale is chosen as $\mu = \pi T$, corresponding to the typical thermal scale of hard excitations in the medium.
This choice avoids an unphysical sensitivity to the infrared region that would arise if the momentum transfer $\sqrt{-t}$ were used as the scale in channels without an intrinsic infrared regulator (such as the $s$- and $u$ channels). At the same time, it preserves the correct asymptotic freedom behavior at high temperatures, ensuring a controlled and consistent implementation of the running coupling.

Consequently, the Debye screening mass $m_D^2 = \frac{N_c}{3} g^2(T) T^2$, as well as the intermediate momentum cutoff scale $-t^\ast \sim m_D^2$, become dynamically dependent on the running coupling. This ensures that the phase-space boundary separating the soft and hard scattering regions evolves consistently with the temperature.


\section{Results and discussions}\label{sec:result}
\begin{figure}[!htbp]
	\begin{center}
		\setlength{\abovecaptionskip}{-0.1mm}
		\includegraphics[width=.47\textwidth]{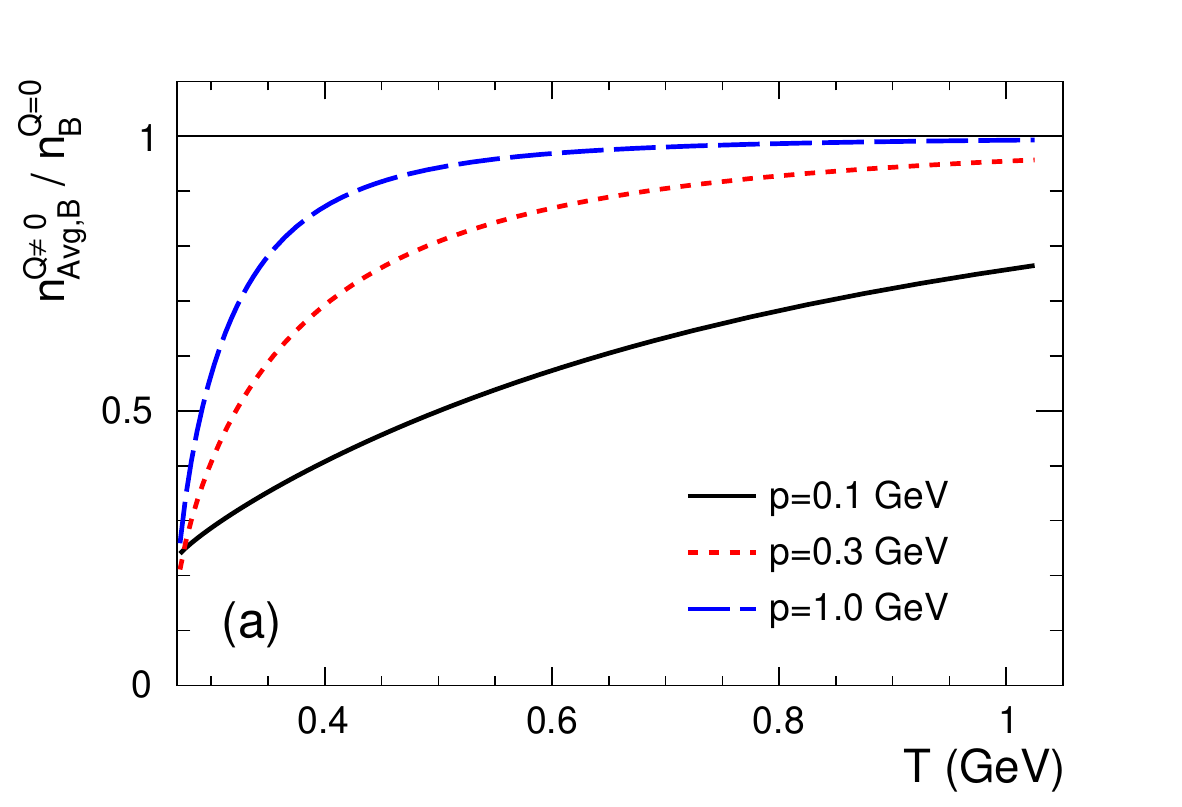}
		\includegraphics[width=.47\textwidth]{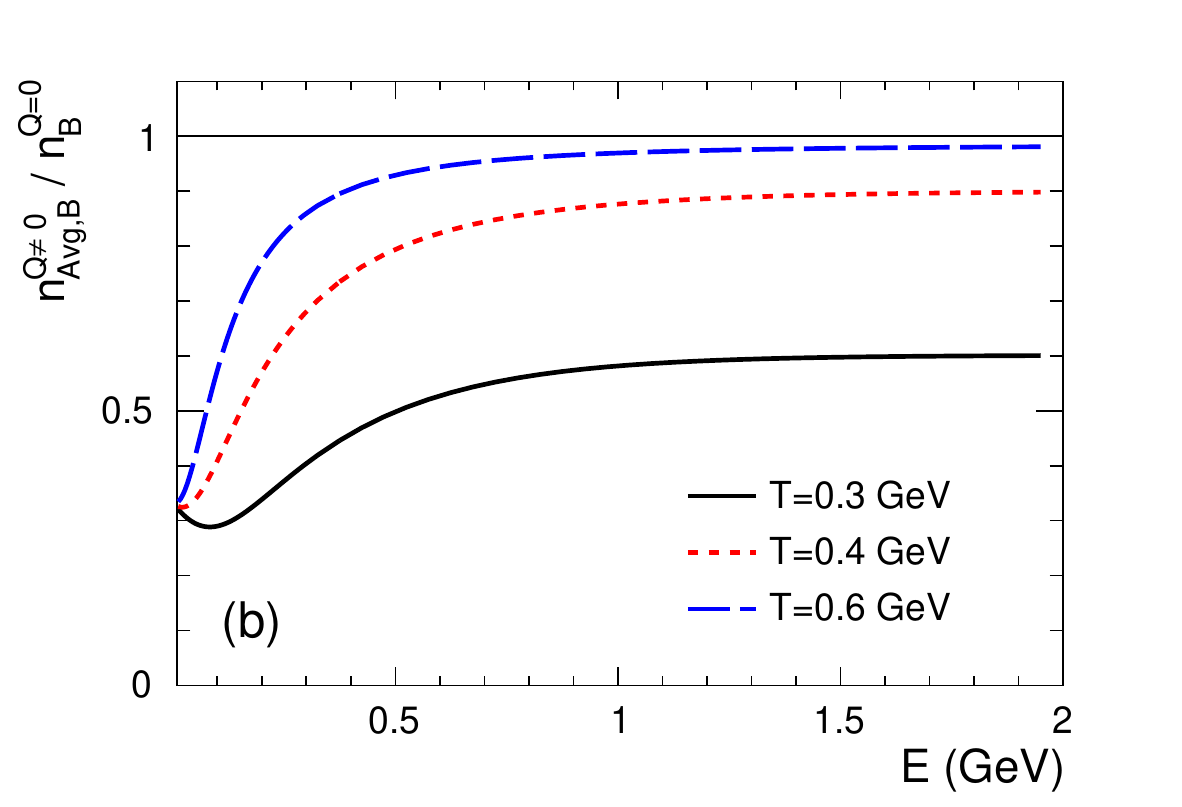}
		\caption{Left (a): comparison of the temperature dependence of the nonperturbative suppression of the bosonic distribution,
		$n_{Avg,B}^{\mathcal{Q}\ne0}/n_{B}^{\mathcal{Q}=0}$,
		for three fixed momentum values: $p=0.1~{\rm GeV}$ (solid black curve),
		$p=0.3~{\rm GeV}$ (dashed red curve) and $p=1.0~{\rm GeV}$ (long-dashed blue curve).
		Right (b): same as panel a but for $n_{Avg,B}^{\mathcal{Q}\ne0}/n_{B}^{\mathcal{Q}=0}$
		as a function of energy for fixed temperatures.} 
		\label{fig:Ratio_nB}
	\end{center}
\end{figure}

Figure~\ref{fig:Ratio_nB} shows the influence of the nonperturbative background field on the thermal distribution of bosonic degrees of freedom,
which is quantified by comparing the color-averaged distribution function,
$n_{Avg,B}^{\mathcal{Q}\ne0}$ [Eqs.~(\ref{eq:Qvector_vsT}), (\ref{eq:PolyaLoop_2}) and (\ref{eq:BFET_ThermalDis_Boson_Avg})],
with the baseline, $n_{B}^{\mathcal{Q}=0}$ [Eq.~(\ref{eq:ThermalDis_Boson})].
The temperature-dependence of the corresponding ratio, $n_{Avg,B}^{\mathcal{Q}\ne0}/n_{B}^{\mathcal{Q}=0}$,
is presented in the panel a for three fixed momenta.
See the legend for details.
The ratio is systematically below unity and exhibits a strong momentum dependence,
with lower-momentum states being more significantly suppressed.
This indicates that the background field preferentially inhibits
the population of low-momentum bosonic modes in the semi-QGP phase.
The monotonic increase of the ratio with temperature, saturating to a stable value at high $T$,
reflects the gradual melting of the nonperturbative background
and the transition toward a perturbative plasma where the suppression is lifted.

This behavior is further elucidated in the panel b of Fig.~\ref{fig:Ratio_nB},
which plots the ratio as a function of energy $E$ for fixed temperatures.
The observed increase of the ratio with energy demonstrates that
the suppression mechanism is most effective for low-energy excitations.
Furthermore, for a given energy, the suppression is markedly stronger at a lower temperature ($T=0.3~{\rm GeV}$)
than at a higher one ($T=0.6~{\rm GeV}$).
These results collectively provide direct evidence that the nonperturbative background field
generates a spectral suppression that is most potent in the low-energy, low-temperature regime of the semi-QGP.
This gapping of the soft thermal spectrum is a key mechanism responsible for the modified transport properties,
such as the suppressed energy loss of heavy quarks reported later,
as it reduces the available density of scatterers (a property of the medium) and soft gluon emission rates.

\begin{figure}[!htbp]
	\begin{center}
		\setlength{\abovecaptionskip}{-0.1mm}
		\includegraphics[width=.47\textwidth]{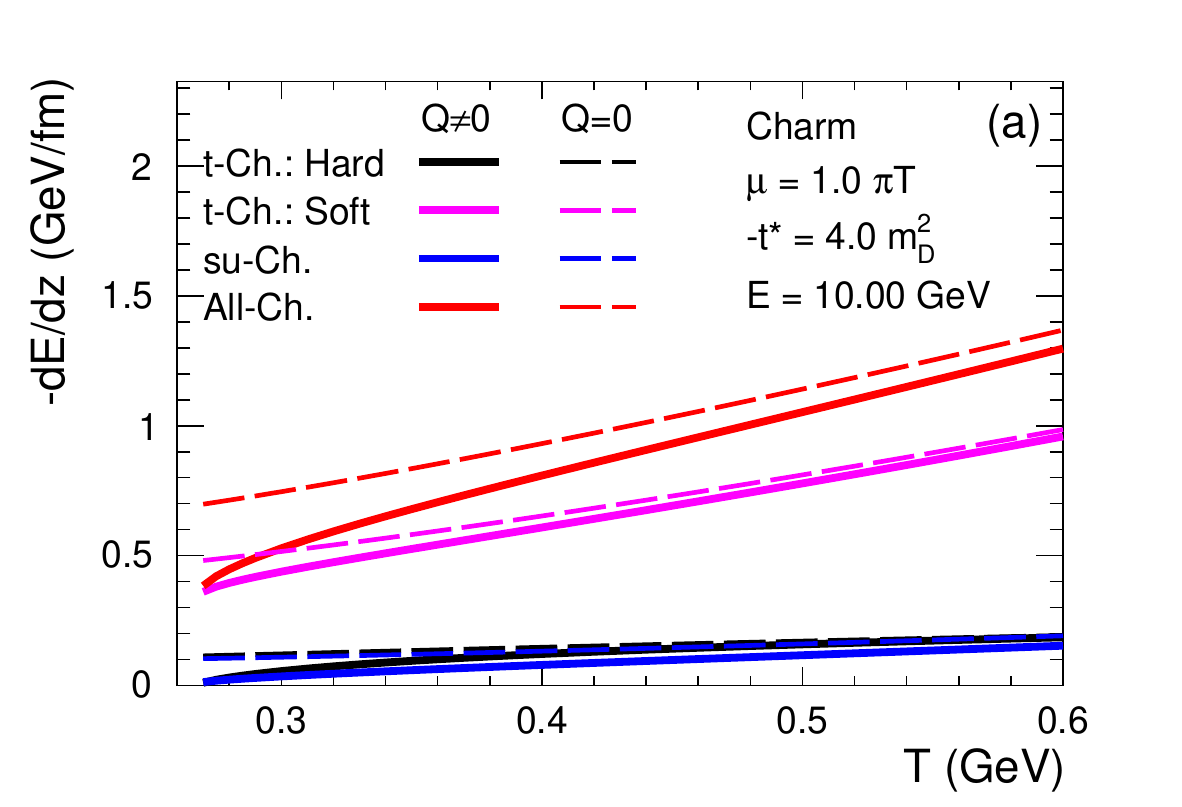}
		\includegraphics[width=.47\textwidth]{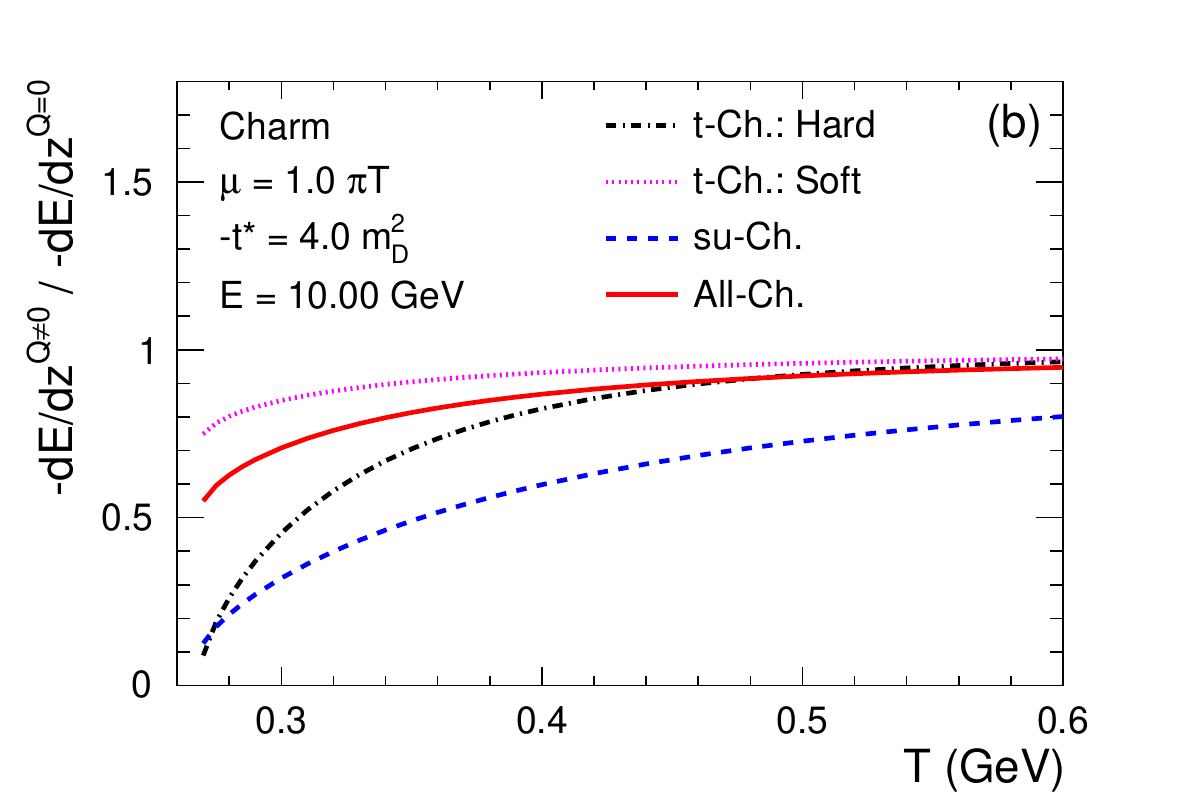}
		\caption{Left (a): comparison of the energy loss $-dE/dz$ as a function of temperature, for a charm quark at a given energy $E=10~{\rm GeV}$, obtained with a nonvanishing background field [$\mathcal{Q}\ne 0$; thick curves; representing the dynamic $Q(T)$ from Eq.~(\ref{eq:Qvector_vsT})] and a vanishing background [$\mathcal{Q}=0$; thin curves; serving strictly as the background-free perturbative baseline]. The contributions from various channels are displayed separately as curves with different styles. Right (b): further comparison of the relative ratio $dE/dz$ between the dynamical $\mathcal{Q}\ne 0$ state and the perturbative $\mathcal{Q}=0$ baseline.}
		\label{fig:dEdz_CmpBkg_vsT}
	\end{center}
\end{figure}
Figure~\ref{fig:dEdz_CmpBkg_vsT} (a) presents the collisional energy loss, $-dE/dz$, for a $10~{\rm GeV}$ charm quark as a function of the medium temperature $T$.
The results are delineated by scattering channel: $t$ channel hard (black), $t$ channel soft (pink), $su$ channel (blue), and the total (red) contributions.
Crucially, across the entire temperature range and for every individual channel, the energy loss in the dynamically running $\mathcal{Q}\neq 0$ scenario is systematically suppressed compared to the $\mathcal{Q}=0$ baseline.
This provides direct evidence that the nonperturbative dynamics of the semi-QGP medium reduce its stopping power against a heavy quark probe.
The suppression is further quantified in the panel b of Fig.~\ref{fig:dEdz_CmpBkg_vsT}, which displays the ratio $(-dE/dz)^{\mathcal{Q}\neq 0} / (-dE/dz)^{\mathcal{Q}=0}$ as a function of $T$.
The overall suppression is most pronounced near the transition region ($T\sim 0.3~{\rm GeV}$) and gradually weakens with increasing temperature, signaling the dissolution of nonperturbative effects.

A detailed channel analysis reveals a clear suppression hierarchy: the $su$ channel is suppressed the most, followed by the $t$ channel hard process, while the $t$ channel soft process is the least affected. This hierarchy reflects their distinct sensitivities to the infrared (IR) phase space:
\begin{enumerate}
	\item[(i)] The $su$ channel (fully exposed binary scattering): With the large heavy quark mass regulating IR divergences~\cite{Peigne:2008nd}, the phase space integrates down to zero momentum ($p_2 \to 0$). This fully exposes the channel to the severe depletion of ultrasoft thermal gluons induced by the semi-QGP background field~\cite{Hidaka:2009hs}, resulting in the most drastic suppression.
	\item[(ii)] The $t$ channel hard process (shielded binary scattering): The requirement of a kinematic cutoff $|t| > |t^\ast| \sim m_D^2$ to regulate Coulomb-like IR divergences acts as a shield. It explicitly prevents the hard scattering from sampling the ultrasoft momentum region where the background field's depletion effect is most severe, leading to a milder suppression.
	\item[(iii)] The $t$ channel soft process (robust collective fields): Governed by momentum transfers $|t| < |t^\ast|$, this process describes interactions with the medium's long-range chromoelectric fields via HTL resummed propagators. Lattice QCD indicates that while free gluon density drops near $T_c$, these macroscopic chromoelectric correlations remain robust~\cite{Bazavov:2018wmo}, making this channel the least affected.
\end{enumerate}

These findings suggest that long-wavelength color interactions provide a smooth dynamical connection between the semi-QGP and perturbative QGP regimes~\cite{Bazavov:2018wmo}.

Replacing the fixed coupling with a 1-loop running coupling $\alpha_s(\pi T)$ naturally reduces the absolute energy loss at high temperatures due to asymptotic freedom. Remarkably, the relative suppression ratio, $(-dE/dz)^{\mathcal{Q}\neq 0} / (-dE/dz)^{\mathcal{Q}=0}$, remains essentially unchanged. This robust consistency confirms that the suppression is fundamentally a kinematic, phase-space effect---driven by the statistical depletion of low-momentum gluons---rather than a dynamic vertex effect. The coupling strength factors out in the ratio, demonstrating the theoretical robustness of the semi-QGP framework.

\begin{figure}[!htbp]
	\begin{center}
		\setlength{\abovecaptionskip}{-0.1mm}
		\includegraphics[width=.47\textwidth]{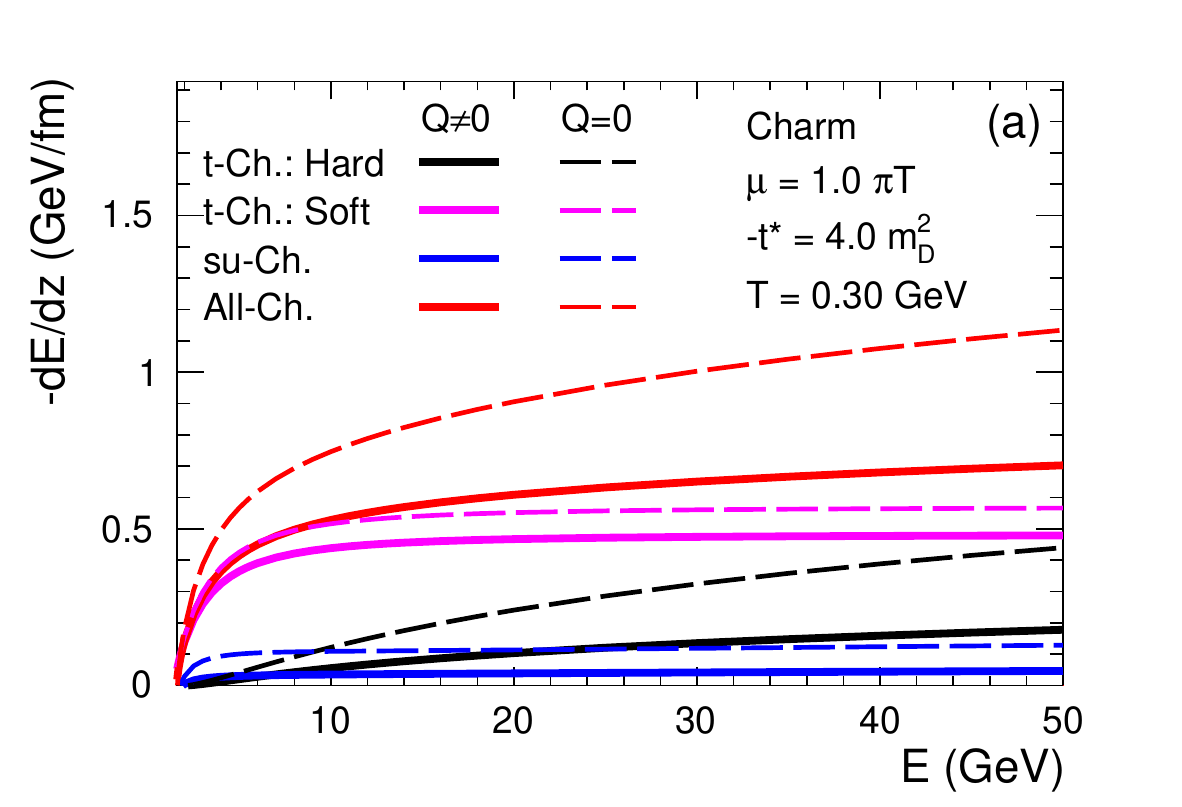}
		\includegraphics[width=.47\textwidth]{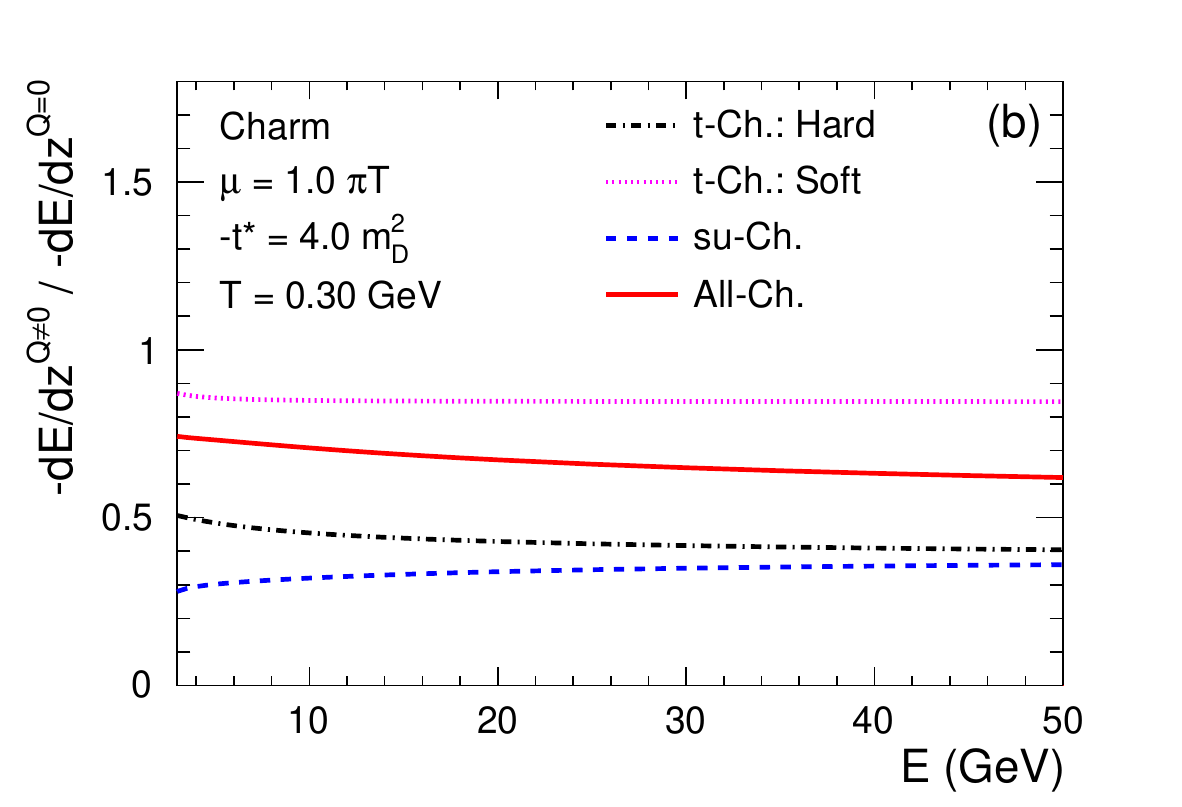}
		\caption{Same as Fig.~\ref{fig:dEdz_CmpBkg_vsT} but as a function of heavy-quark energy at fixed temperature $T=0.3~{\rm GeV}$. Thick curves represent the dynamic $\mathcal{Q}\ne 0$ state, while thin curves represent the $\mathcal{Q}=0$ baseline.}
		\label{fig:dEdz_CmpBkg_vsE}
	\end{center}
\end{figure}
Figure~\ref{fig:dEdz_CmpBkg_vsE} shows the energy dependence of the energy loss. We find that the background-field induced suppression of collisional energy loss exhibits only a weak dependence on the incident heavy-quark energy.
This is a natural consequence of how the background modifies the medium:
the Polyakov-loop–like background effectively reduces the occupation numbers
of colored excitations and therefore multiplies the scattering rates by a
temperature-dependent factor that reflects incomplete color liberation.
Because this suppression primarily alters the availability of scatterers
rather than the kinematics of the probe,
it acts almost multiplicatively on the various elastic cross sections
and is to leading order independent of the heavy-quark energy.

A more dynamical explanation can be gleaned from the channel decomposition.
In the soft part, the $t$ channel dominated regime--responsible for the bulk of collisional energy loss near $T_c$--the probe-energy dependence
of $-dE/dx$ is intrinsically weak (at most logarithmic from the perturbative calculation in the high-energy limit~\cite{Peng:2024zvf}),
because scattering is governed by long-wavelength gluon exchanges regulated by the Debye mass.
The same qualitative behavior persists for the hard $t$ channel and $su$ channel processes:
although they involve larger momentum transfer or different kinematic topologies,
their partial energy losses per scattering remain nearly constant for relativistic heavy quarks,
with only mild logarithmic corrections.
Since the background field primarily suppresses the color occupation factors
rather than modifying the scattering kinematics,
it reduces all these channels by an approximately energy-independent factor.
Consequently, the overall background-induced suppression of the collisional energy loss
shows only a weak dependence on the incident quark energy,
reflecting that the semi-QGP primarily alters the density and color activity of the medium
while leaving the energy scaling of microscopic scattering largely intact.

\begin{figure}[!htbp]
\begin{center}
\setlength{\abovecaptionskip}{-0.1mm}
\includegraphics[width=.47\textwidth]{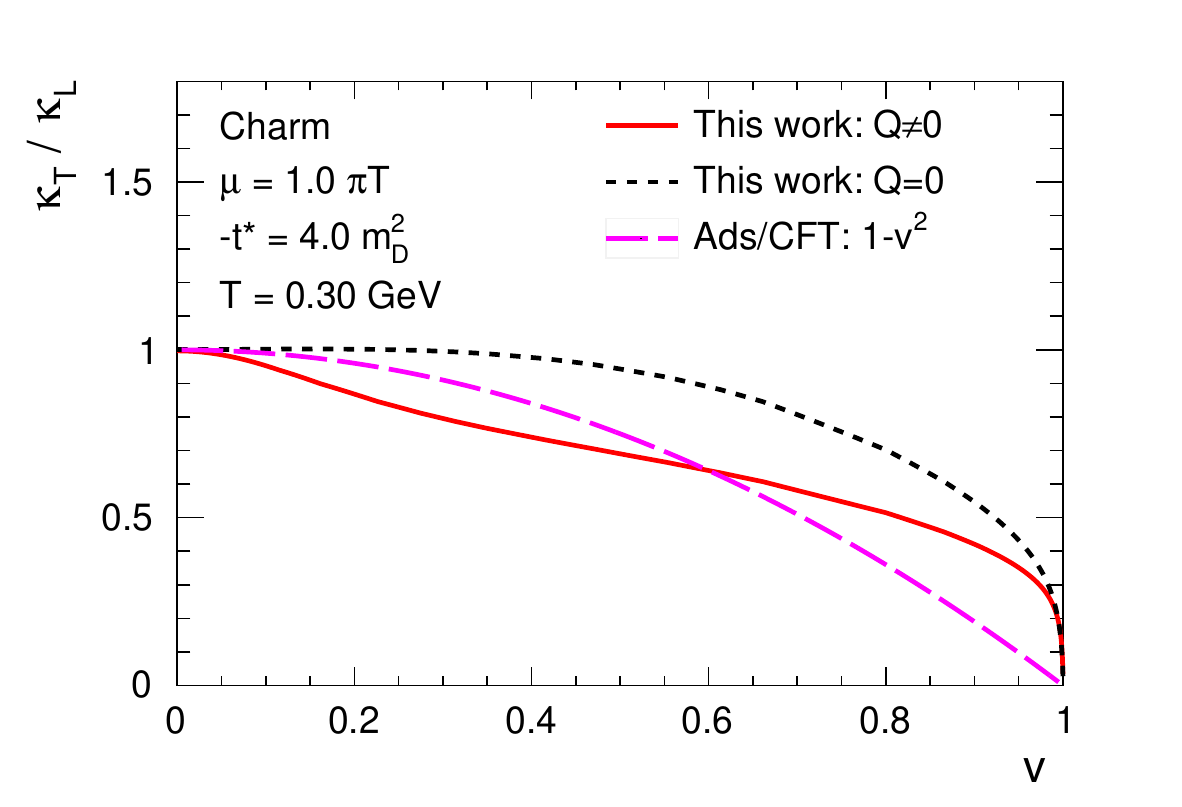}
\caption{(The ratio $R_{TL}=\kappa_{T}/\kappa_{L}$ of the transverse to
longitudinal momentum diffusion coefficients as a function of the heavy-quark velocity $v$.
The result for the background-modified case ($\mathcal{Q}\ne0$) is shown as a solid red curve,
while that including the background-free effects ($\mathcal{Q}=0$) is shown as dashed black curve.
The result from the AdS/CFT~\cite{Gubser:2006bz, Gubser:2006nz, Casalderrey-Solana:2007ahi}
is shown as long-dashed pink curve for comparison.}
\label{fig:Ratio_KappaTKappaL_vsT}
\end{center}
\end{figure}
Figure~\ref{fig:Ratio_KappaTKappaL_vsT} displays the ratio $R_{TL}=\kappa_T/\kappa_L$ as a function of the heavy-quark velocity $v$ for both vanishing ($\mathcal{Q}=0$) and nonvanishing ($\mathcal{Q}\neq0$) background color fields.
In the static limit $v\to0$ the ratio approaches unity, $R_{TL}\to1$,
consistent with isotropic momentum diffusion when longitudinal and transverse fluctuations are indistinguishable.
As the velocity increases $R_{TL}$ decreases monotonically in both cases,
reflecting that kinematic factors and the redistribution of energy-momentum transfers
make longitudinal broadening increasingly efficient relative to transverse broadening for a fast-moving probe.
This anisotropy can be intuitively understood in terms of the Lorentz-contracted correlation length of medium fields
in the quark rest frame and the kinematic enhancement of energy-transfer modes along the direction of motion;
together these effects shift the balance of scattering phase space
toward fluctuations that contribute more strongly to $\kappa_L$ than to $\kappa_T$.
Our small-$v$ behavior agrees with previous perturbative and lattice-QCD studies~\cite{Moore:2004tg, Caron-Huot:2007rwy, Alberico:2013bza, Banerjee:2012ra, Brambilla:2020siz}.

When a nonzero background field is present ($\mathcal{Q}\neq0$)
the ratio $R_{TL}$ is systematically suppressed across the velocity range relative to the perturbative baseline ($\mathcal{Q}=0$).
This suppression should be interpreted as a relative enhancement of the longitudinal contribution
rather than a direct increase of the microscopic longitudinal coupling:
the Polyakov-loop--like background primarily reduces the occupation numbers of colored quasiparticles
and depletes the hard/finite-momentum processes that contribute disproportionately to transverse momentum broadening.
By contrast, the longitudinal diffusion retains a larger share of contributions from soft,
long-wavelength chromoelectric correlations whose infrared structure is less affected by the background.
The net result is a stronger fractional reduction of $\kappa_T$ than of $\kappa_L$,
and consequently a reduced $R_{TL}$.
While this qualitative trend bears similarity to
holographic findings [for example $R_{TL}^{\rm AdS/CFT}(v)\simeq 1-v^{2}$ in certain strongly coupled models~\cite{Gubser:2006bz, Gubser:2006nz, Casalderrey-Solana:2007ahi}]
the underlying microscopic mechanisms differ:
in our semi-QGP framework the effect originates from occupancy and channel reweighting
in a background-modified medium rather than from intrinsically strong-coupling dynamics.
	
\begin{figure}[!htbp]
	\begin{center}
		\setlength{\abovecaptionskip}{-0.1mm}
		\includegraphics[width=.47\textwidth]{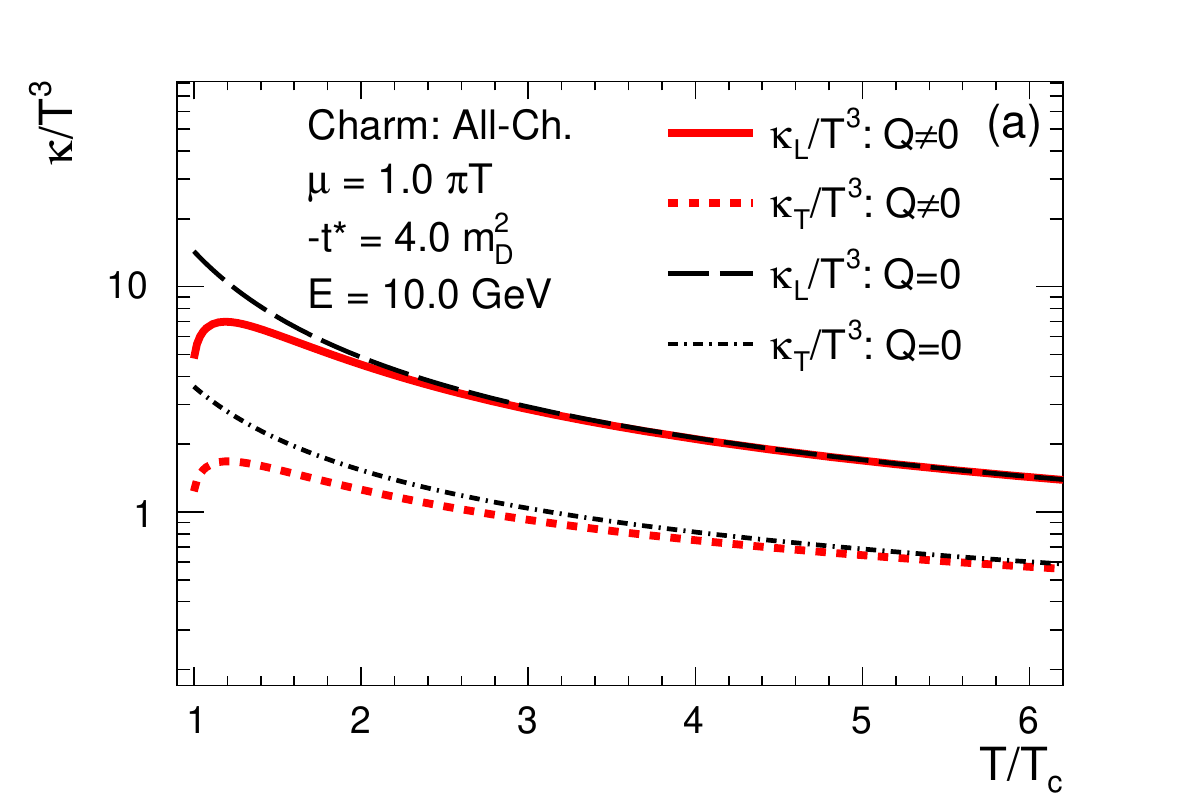} 
		\includegraphics[width=.47\textwidth]{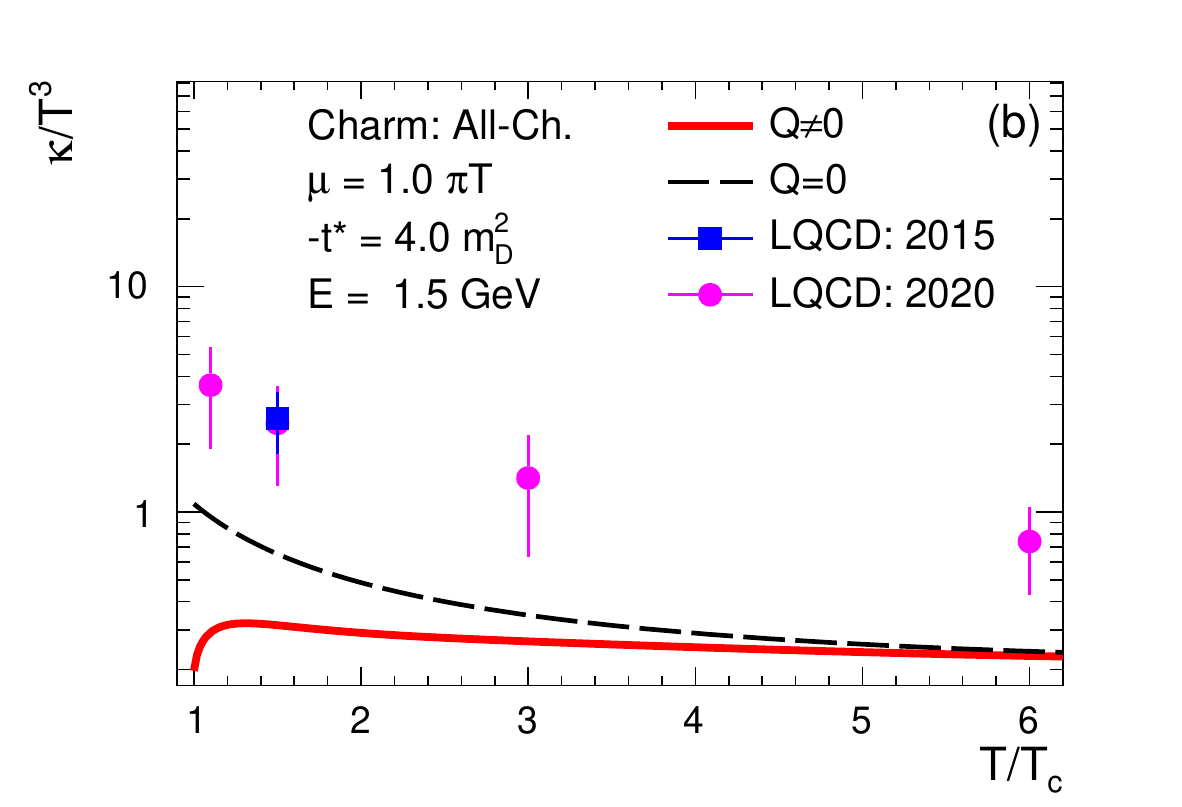}
		\caption{(Color online) The dimensionless momentum diffusion coefficients $\kappa_{T/L}/T^3$ as functions of the scaled temperature $T/T_c$. Panel (a) shows the results for a high-energy charm quark ($E=10~{\rm GeV}$), comparing the semi-QGP framework ($\mathcal{Q} \ne 0$; thick lines) with the background-free perturbative baseline ($\mathcal{Q} = 0$; thin lines). Panel (b) presents the near-static limit ($E=1.5~{\rm GeV}$), where $\kappa_L \approx \kappa_T \approx \kappa$. In this panel, the theoretical evaluations are directly compared with pure-glue ($N_f=0$) Lattice QCD data~\cite{Brambilla:2020siz, Francis:2015daa} to highlight the mechanistic differences near the phase transition.}
		\label{fig:Kappa_over_T3}
	\end{center}
\end{figure}
Figure~\ref{fig:Kappa_over_T3} shows the dimensionless momentum diffusion coefficients, $\kappa_{T/L}/T^3$, as functions of the scaled temperature $T/T_c$ for two representative kinematic regimes. For a high-energy charm quark ($E=10~{\rm GeV}$) in Fig.~\ref{fig:Kappa_over_T3}(a), both transverse and longitudinal diffusion coefficients exhibit a strong suppression in the presence of the nonperturbative background field ($\mathcal{Q} \neq 0$), particularly near the pseudocritical region $T \approx T_c$. In the perturbative baseline ($\mathcal{Q}=0$), the rapid growth of the running coupling near $T_c$ artificially inflates $\kappa_{T/L}/T^3$, signaling a breakdown of the weak-coupling expansion. The semi-QGP framework naturally regulates this divergence by depleting the soft gluonic degrees of freedom via the Polyakov loop, thereby controlling the effective scattering phase space. As the temperature increases and the background field melts, the $\mathcal{Q} \neq 0$ results smoothly converge to the perturbative baseline, reflecting the restoration of a weakly coupled plasma.

In the near-static limit ($E \approx m_c = 1.5~{\rm GeV}$, $v\to 0$) shown in Fig.~\ref{fig:Kappa_over_T3}(b), the diffusion becomes isotropic ($\kappa_L \approx \kappa_T \approx \kappa$). To isolate the nonperturbative medium effects, we directly compare our pure-gauge theoretical evaluations with pure-glue ($N_f=0$) lattice QCD data. Both approaches exhibit a decreasing trend of $\kappa/T^3$ with increasing temperature. However, a significant quantitative discrepancy remains: our evaluation ($\kappa/T^3 \lesssim 0.32$) is substantially smaller than lattice estimates ($\kappa/T^3 \sim 0.4 - 5.4$). This discrepancy explicitly isolates the contribution of strongly coupled gluonic dynamics. Near the phase transition ($T \approx T_c$), this large gap is primarily driven by nonperturbative chromomagnetic interactions and spatial correlations (e.g., chromomagnetic monopoles), which dominate transport properties but are intrinsically absent in our leading-order (LO) chromoelectric framework. Furthermore, the discrepancy persists at higher temperatures (e.g., $T \sim 6T_c$) due to the poor convergence of the thermal perturbative expansion---where purely gluonic NLO corrections are exceptionally large---combined with residual nonperturbative magnetic screening effects. Consequently, our framework establishes a rigorous, purely perturbative lower baseline for the chromoelectric sector, strictly delineating the boundary where nonperturbative magnetic contributions and higher-order resummations must be integrated.

	
\section{summary}\label{sec:summary}

We have extended the SHFM framework by incorporating a temperature-dependent background field
that accounts for nonperturbative QCD effects near $T_c$.
This unified approach allows for a continuous interpolation between the perturbative and quasiconfining regimes
and is valid in both the $E_{Q}\gg m_{Q}^{2}/T$ (small momentum transfer)
and $E_{Q}\ll m_{Q}^{2}/T$ (large momentum transfer) limits.
The computed energy loss and diffusion coefficients show clear suppression,
especially near the critical temperature,
due to reduced screening and modified thermal distribution functions,
emphasizing the significance of nonperturbative color correlations in this temperature regime.

We end with discussions on a few important caveats in the present study
that deserve emphasis and that call for future investigations:
\begin{itemize}
	\item[i.]
	In this work (and in many other studies such as Refs.~\cite{Gupta:2017gbs, Berges:2020fwq, Du:2024riq}),
	we focus exclusively on a gluonic plasma as a first step toward a comprehensive QCD analysis.
	The contributions from thermal quarks are omitted,
	although they play important roles in medium screening and heavy-quark scattering.
	Incorporating these effects consistently within an effective theory is highly nontrivial,
	as it requires accounting for fermionic Matsubara modes and nonperturbative quark dynamics near $T_c$.
	Focusing on the gluonic sector thus provides a tractable framework that
	captures the dominant features of the medium and
	serves as a baseline for future extensions including thermal-quark contributions.
	\item[ii.]
	It may be noted that we have not yet incorporated the inelastic radiative processes
	associated with medium-induced gluon emission and absorption.
	These processes become increasingly important at high momentum,
	where radiative energy loss dominates over elastic scatterings~\cite{HQQGPRapp10, Li:2019wri, He:2022ywp}.
	Their consistent inclusion requires extending the current approach to
	account for the Landau-Pomeranchuk-Migdal effect,
	arising from the non-Abelian interference among multiple scatterings.
	While implementing these effects within the same effective-theory framework remains challenging,
	it is essential for a quantitative description of heavy-flavor energy loss and
	nuclear modification factors at high transverse momentum at RHIC and the LHC.
	\item[iii.]
	The momentum diffusion coefficients obtained in this study serve as key inputs for phenomenological applications,
	particularly within our heavy-quark transport modeling
	framework (Langevin-transport with gluon radiation, LGR)~\cite{Li:2019lex, Li:2018izm, Li:2018jba, Li:2020umn}.
	These coefficients characterize the microscopic interactions between heavy quarks and the thermal medium,
	encoding the strength of both transverse and longitudinal momentum broadening.
	Coupling the present results to the LGR framework would be a natural and promising extension,
	as it enables a direct, quantitative comparison with experimental heavy-flavor observables
	across the full momentum range at RHIC and the LHC.
	Such a unified treatment would bridge the connection between the microscopic transport properties
	derived here and the macroscopic phenomenology of heavy-quark dynamics in hot gluon matter.
\end{itemize}

\begin{acknowledgements}
	The authors are grateful to Yun Guo and Shu Lin for helpful discussions and communications.
	This work is supported by the National Natural Science Foundation of China (NSFC) under Grants No. 12375137 and No. 12005114.
\end{acknowledgements}



\appendix

\section{Calculation of the soft interaction rate in a background field}\label{appendix:Gamma_Soft_BFET}
\setcounter{equation}{0}
\renewcommand\theequation{A\arabic{equation}}

In this Appendix, we calculate the soft interaction rates for the off-diagonal and diagonal gluons.

\begin{figure}[!htbp]
	\centering
	\includegraphics[width=.45\textwidth]{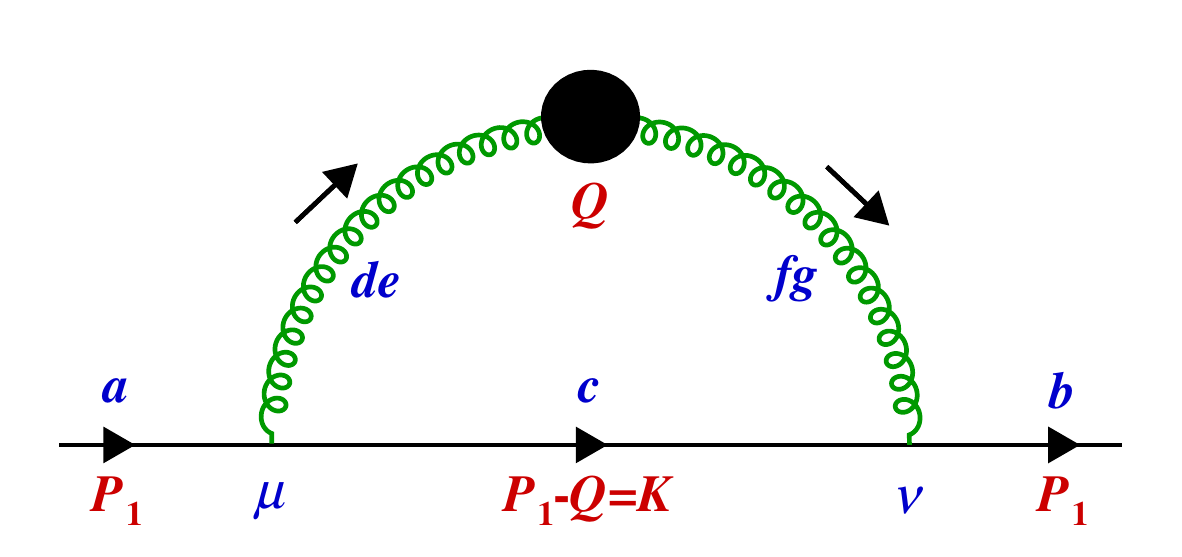}
	\caption{Feynman diagram for the quark self-energy with the gluon HTL-resummed propagator (bolb) in the double line basis.}
	\label{fig:Quark_SelfEnergy_BFET}
\end{figure}
Figure~\ref{fig:Quark_SelfEnergy_BFET} shows the diagram for the heavy quark self-energy,
where the blob on the gluon line represents an effective gluon propagator $D^{de,fg}_{\mu\nu}$.
In this case, the effective self-energy for the heavy quark is given by
\begin{equation}
	\begin{aligned}
		\Sigma_{ab}^{de,fg} &= -ig^{2}\delta_{ab} \sum_{c,d,e,f,g=1}^{N_{c}} \bigr(t^{de}_{ac}t^{fg}_{cb}\bigr)
		\int\frac{d^{4}Q}{(2\pi)^{4}}D_{\mu\nu}^{de,fg}\gamma^{\mu}
		\frac{1}{\slashed{K}^{c} -m_{1}} \gamma^{\nu},
	\end{aligned}
\end{equation}
where, $t^{de}_{ac}$ and $t^{fg}_{cb}$ are the generators in double line basis;
$\sum_{colors}$ is the summation over color indices;
$D_{\mu\nu}^{de,fg}$ is the HTL-resummed gluon propagator in the background field
[Eqs.~(\ref{eq:BFET_GluonPropagator_OffDiag_T}) and (\ref{eq:BFET_GluonPropagator_OffDiag_L}) for its transverse and longitudinal components].
$\slashed{K}^{c}=\slashed{P}_{1}^{a}-\slashed{Q}^{de}$ is the four-momentum of the internal quark,
which satisfies momentum conservation as usual.
$Q^{de}=(q_{0}^{de},\vec{q}\;)=(q_{0}+i\mathcal{Q}^{de},\vec{q}\;)$
is the four-momentum of the HTL-resummed propagator with color index $de$ in the adjoint representation,
with $\mathcal{Q}^{de}=\mathcal{Q}^{d}-\mathcal{Q}^{e}$ the color vector.

We now evaluate the trace in Eq.~(\ref{eq:Gamma_BFET_Soft}) with the off-diagonal gluon propagators, yielding
\begin{equation}
	\begin{aligned} \label{eq:BFET_Trace1}
		&Tr [(\slashed{P}_{1}^{a}+m_{1}) \cdot \Sigma_{ab}^{de,fg}]_{off-diag} =
		4g^{2}T\delta_{ab} \sum_{\substack{c,d,e,f,g=1 \\ (d\ne e)}}^{N_{c}} \bigr(t^{de}_{ac}t^{fg}_{cb}\bigr) \sum_{q_{0}} \int_q \frac{1}{(k^{c}_{0})^{2}-E_{k}^{2}} \\
		&\biggr\{ \Delta_{L,off-diag}^{de,fg} \bigr[ (p_{1;0}^{a})^{2} - p_{1;0}^{a}q_{0}^{de} + \vec{p}_{1}^{\;2} - \vec{p}_{1}\cdot\vec{q} + m_{1}^{2} \bigr]
		+
		2\Delta_{T,off-diag}^{de,fg} \bigr[ (p_{1;0}^{a})^{2} - p_{1;0}^{a}q_{0}^{de} - (\vec{p}_{1}\cdot\widehat{q})^{2} + \vec{p}_{1}\cdot\vec{q} - m_{1}^{2} \bigr] \biggr\},
	\end{aligned}
\end{equation}
where $E_{k}=\sqrt{\vec{k}^{\;2}+m_{1}^{2}}$ is the energy of internal quark;
$\widehat{q}=\vec{q}/|\vec{q}\;|$ is the unit vector along $\vec{q}$.

The mixed representation of the effective gluon propagators is given by
\begin{equation}
	\begin{aligned} \label{eq:BFET_GluonPropagator_MixRep}
		\Delta_{T/L,off-diag}^{de,fg}(q_{0},\vec{q}\;) = \frac{1}{2\pi} \int^{\beta}_{0} d\tau_{1} e^{q^{de}_{0}\tau_{1}}
		\int^{+\infty}_{-\infty} d\omega \rho_{T/L,off-diag}^{de,fg} e^{-\omega\tau_{1}} \bar{n}_{B}(\omega-i\mathcal{Q}^{de}),
	\end{aligned}
\end{equation}
with $q=|\vec{q}\;|$; the spectral density functions are defined
by the imaginary part of the retarded propagator,
\begin{equation}
	\begin{aligned} \label{eq:BFET_GammaRhoTL1}
		\rho_{T/L,off-diag}^{de,fg}(\omega,q)=2\cdot Im \Delta_{T/L,off-diag}^{de,fg} (\omega+i\eta,q),
	\end{aligned}
\end{equation}
which is given by setting $q_{0}^{de}= \omega+i\eta \; (\eta\to 0^{+})$ in
Eqs.~(\ref{eq:BFET_GluonPropagator_OffDiag_T}) and (\ref{eq:BFET_GluonPropagator_OffDiag_L}).
It gives
\begin{subequations}
	\begin{align}
		\rho_{T,off-diag}^{de,fg}(\omega,q) =& \bigr(\delta^{dg}\delta^{ef}\bigr) \frac{\pi \omega (\mathcal{M}^{2}_{D})^{de}_{off-diag}}{2q^{3}} (q^{2}-\omega^{2}) \Biggr\{ \biggr[ q^{2}-\omega^{2} +  \nonumber \\
		& \frac{\omega^{2} (\mathcal{M}^{2}_{D})^{de}_{off-diag}}{2q^{2}} \bigr(1+\frac{q^{2}-\omega^{2}}{2\omega q} ln\frac{q+\omega}{q-\omega}\bigr) \biggr]^{2}
		+ \biggr[ \frac{\pi \omega (\mathcal{M}^{2}_{D})^{de}_{off-diag}}{4q^{3}} (q^{2}-\omega^{2}) \biggr]^{2} \Biggr\}^{-1}, \label{eq:BFET_GammaRhoT_Soft_OffDiag}
		\\
		\rho_{L,off-diag}^{de,fg}(\omega,q) =& \bigr(\delta^{dg}\delta^{ef}\bigr) \frac{\pi \omega (\mathcal{M}^{2}_{D})^{de}_{off-diag}}{q}
		\Biggr\{ \biggr[ q^{2}+ (\mathcal{M}^{2}_{D})^{de}_{off-diag}\bigr(1-\frac{\omega}{2q} ln\frac{q+\omega}{q-\omega}\bigr) \biggr]^{2} + \nonumber \\
		& \biggr[ \frac{\pi \omega (\mathcal{M}^{2}_{D})^{de}_{off-diag}}{2q} \biggr]^{2} \Biggr\}^{-1}, \label{eq:BFET_GammaRhoL_Soft_OffDiag}
	\end{align}
\end{subequations}
with $(\mathcal{M}^{2}_{D})^{de}_{off-diag}$ the $\mathcal{Q}$-modified
Debye screening mass squared [Eq.~(\ref{eq:BFET_OffDiag_DebMas})].

For quark propagator
\begin{equation}
	\begin{aligned} \label{eq:BFET_QuarkPropagator_MixRep}
		\frac{1}{(k_{0}^{c})^{2}-E_{k}^{2}} &= -\frac{1}{2E_{k}} \int^\beta_{0} d\tau_{2} e^{k^{c}_{0} \tau_{2}}
		\biggr[ e^{-E_{k}\tau_{2}} \bar{n}_{F}(E_{k}-i\mathcal{Q}^{c}) - e^{E_{k}\tau_{2}} n_{F}(E_{k}+i\mathcal{Q}^{c}) \biggr],
	\end{aligned}
\end{equation}
with $k_{0}^{c}=p_{1;0}^{a}-q_{0}^{de}$.
Inserting Eqs.~(\ref{eq:BFET_GluonPropagator_MixRep}) and (\ref{eq:BFET_QuarkPropagator_MixRep}) into Eq.~(\ref{eq:BFET_Trace1}),
we have
\begin{equation}
	\begin{aligned} \label{eq:BFET_Trace2}
		&Tr [(\slashed{P}_{1}^{a}+m_{1}) \cdot \Sigma_{ab}^{de,fg}]_{off-diag} =
		-\frac{g^{2}T}{\pi}\delta_{ab} \sum_{\substack{c,d,e,f,g=1 \\ (d\ne e)}}^{N_{c}} \bigr(t^{de}_{ac}t^{fg}_{cb}\bigr)
		\int_{q}\frac{1}{E_{k}} \int_{0}^{\beta}d\tau_{1} \int_{0}^{\beta}d\tau_{2}
		\biggr[ \sum_{q_{0}}e^{q_{0}^{de}(\tau_{1}-\tau_{2})} \biggr] \\
		& e^{p_{1;0}^{a}\tau_{2}} \biggr[e^{-E_{k}\tau_{2}} \bar{n}_{F}(E_{k}-i\mathcal{Q}^{c}) - e^{E_{k}\tau_{2}} n_{F}(E_{k}+i\mathcal{Q}^{c})\biggr] \\
		& \int^{+\infty}_{-\infty}d\omega e^{-\omega\tau_{1}} \bar{n}_{B}(\omega-i\mathcal{Q}^{de})
		\biggr\{\rho_{L,off-diag}^{de,fg} \bigr[ (p_{1;0}^{a})^{2} - p_{1;0}^{a}q_{0}^{de} + \vec{p}_{1}^{\;2} - \vec{p}_{1}\cdot\vec{q} + m_{1}^{2} \bigr] + \\\
		& 2\rho_{T,off-diag}^{de,fg} \bigr[ (p_{1;0}^{a})^{2} - p_{1;0}^{a}q_{0}^{de} - (\vec{p}_{1}\cdot\widehat{q})^{2} + \vec{p}_{1}\cdot\vec{q} - m_{1}^{2} \bigr] \biggr\}.
	\end{aligned}
\end{equation}
Using the identity,
\begin{equation}
	\sum_{q_{0}}e^{q_{0}(\tau_{1}-\tau_{2})} = \beta\delta(\tau_{1}-\tau_{2}),
\end{equation}
and its derivative, we can carry out the sum over $q_{0}$,
and then evaluate the $\tau_{1}$ and $\tau_{2}$ integrals in Eq.~(\ref{eq:BFET_Trace2}).
As introduced in Ref.~\cite{PhysRevD.44.1298},
to eliminate the exponential of $p_{1;0}$,
we can set $e^{\beta p_{1;0}}=-1$ as $p_{1;0}=i(2n+1)\pi T$.
In this case the heavy quark energy can be analytically continued to the real Minkowski energy
$p_{1;0}^{a}=E_{1}+i\epsilon$ required in Eq.~(\ref{eq:Gamma_BFET_Soft}).
It yields
\begin{equation}
	\begin{aligned} \label{eq:BFET_Trace3}
		Tr [(\slashed{P}_{1}^{a}+m_{1}) \cdot \Sigma_{ab}^{de,fg}]_{off-diag} =&
		\frac{2g^{2}}{\pi}\delta_{ab} \sum_{\substack{c,d,e,f,g=1 \\ (d\ne e)}}^{N_{c}} \bigr(t^{de}_{ac}t^{fg}_{cb}\bigr)
		\int_{q}\frac{E_{1}^{2}}{E_{k}} \int^{+\infty}_{-\infty}d\omega \bar{n}_{B}(\omega-i\mathcal{Q}^{de}) \\
		& \biggr[ \frac{\bar{n}_{F}(E_{k}-i\mathcal{Q}^{c})}{E_{1}-E_{k}-\omega+i\epsilon} - \frac{n_{F}(E_{k}+i\mathcal{Q}^{c}) }{E_{1}+E_{k}-\omega+i\epsilon} \biggr] \bar{n}_{F}^{-1}(E_{1}-i\mathcal{Q}^{a}) \\
		& \biggr\{\rho_{L,off-diag}^{de,fg} \Bigr( 1-\frac{\omega+\vec{v}\cdot\vec{q}}{2E_{1}} \Bigr) +
		\rho_{T,off-diag}^{de,fg} \Bigr[ \vec{v}_{1}^{\;2}(1-(\widehat{v}\cdot\widehat{q})^{2}-\frac{\omega-\vec{v}\cdot\vec{q}}{E_{1}}) \Bigr] \biggr\}.
	\end{aligned}
\end{equation}
For consistency, we take the same approximations as for $\mathcal{Q}=0$:
(i) since heavy quark kinematic energy is much larger than the underlying medium temperature $E_{1/k}\gg T\sim\mathcal{Q}$,
$n_{F}$ is exponentially suppressed and can be dropped;
(ii) as we are interested in the low $\omega$ region $\omega\sim T\ll E_{1}\sim E_{k}$,
the term involving $1/(E_{1}+E_{k}-\omega+i\epsilon)$ vanishes;
(iii) the kinematical approximation $E_{k}\approx E_{1}-\vec{v} \cdot \vec{q}$ is valid for collinear processes.
Consequently, Eq.~(\ref{eq:BFET_Trace3}) can be further simplified as,
\begin{equation}
	\begin{aligned} \label{eq:BFET_Trace4}
		Tr [(\slashed{P}_{1}^{a}+m_{1}) \cdot \Sigma_{ab}^{de,fg}]_{off-diag} =&
		\frac{2g^{2}E_{1}}{\pi}\delta_{ab} \sum_{\substack{c,d,e,f,g=1 \\ (d\ne e)}}^{N_{c}} \bigr(t^{de}_{ac}t^{fg}_{cb}\bigr)
		\int_{q} \int^{+\infty}_{-\infty}d\omega \biggr\{\bar{n}_{B}(\omega-i\mathcal{Q}^{de}) \cdot \frac{1}{E_{1}-E_{k}-\omega+i\epsilon} \biggr\} \\		
		& \biggr\{\rho_{L,off-diag}^{de,fg} +
		\rho_{T,off-diag}^{de,fg} \Bigr[ \vec{v}_{1}^{\;2}(1-(\widehat{v}\cdot\widehat{q})^{2}) \Bigr] \biggr\}.
	\end{aligned}
\end{equation}

When computing the imaginary part of loop integrals in real-time thermal field theory,
it is often necessary to evaluate expressions involving a product of a thermal distribution function
and a propagator with an infinitesimal imaginary part $i\epsilon$, as shown in Eq.~(\ref{eq:BFET_Trace4}).
A standard and well-motivated approximation is to retain only the contribution from the propagator’s pole,
while replacing the accompanying smooth prefactor with its real part~\cite{Kapusta:2006pm, Bellac:2011kqa}:
\begin{equation}
	\begin{aligned} \label{eq:Imag_Approx1}
		Im \biggr\{ \bar{n}_{B}(\omega-i\mathcal{Q}^{de}) \cdot \frac{1}{E_{1}-E_{k}-\omega+i\epsilon} \biggr\}
		\approx &  Re \bigr[ \bar{n}_{B}(\omega-i\mathcal{Q}^{de}) \bigr] \cdot Im \bigr[ \frac{1}{E_{1}-E_{k}-\omega+i\epsilon} \bigr] \\
		\approx & -\pi \cdot \mathcal{N}(\omega,\mathcal{Q}^{de}) \cdot \delta(\omega-\vec{v}\cdot\vec{q}\;),
	\end{aligned}
\end{equation}
with
\begin{equation}
	\begin{aligned} \label{eq:Imag_Approx2}	
		\mathcal{N}(\omega,\mathcal{Q}^{de}) =& \frac{1-e^{-\beta\omega}cos(\beta\mathcal{Q}^{de})}{\bigr[e^{-\beta\omega}-cos(\beta\mathcal{Q}^{de})\bigr]^{2}+sin^{2}(\beta\mathcal{Q}^{de})}.
	\end{aligned}
\end{equation}
We note that $\mathcal{N}(\omega,\mathcal{Q}^{de}=0)=\bar{n}_{B}(\omega)$,
which will be used for the diagonal gluons [Eq.~(\ref{eq:Gamma_ColorAvg_BFET_Soft_Diag})].

The approximation, as shown in Eq.~(\ref{eq:Imag_Approx1}), is justified both mathematically and physically.
Mathematically, the imaginary part of the propagator $1/(E_{1}-E_{k}-\omega+i\epsilon)$ yields a sharply peaked delta function
due to the presence of a simple pole $\omega=E_{1}-E_{k}$ near the real axis,
whereas the distribution function $\bar{n}_{B}(\omega-i\mathcal{Q}^{de})$
is analytic and varies slowly across the energy domain.
Physically, the delta function encodes the on-shell condition for particle propagation or decay,
while the distribution function serves as a statistical weight that modulates
but does not generate any singular behavior.
Therefore, neglecting the subleading imaginary part of the distribution function
introduces no significant error in the extraction of physical observables
such as damping rates or spectral widths.
This approach is consistent with the Cutkosky cutting rules and
the optical theorem in thermal field theory,
and has been widely employed in seminal works such as
Weldon’s analysis of self-energy discontinuities at finite temperature~\cite{WeldonPRD83}.

With Eq.~(\ref{eq:Imag_Approx1}) we can calculate the imaginary part of the trace [Eq.~(\ref{eq:BFET_Trace4})],
and get the corresponding scattering rate [Eq.~(\ref{eq:Gamma_BFET_Soft})]
\begin{equation}
	\begin{aligned} \label{eq:Gamma_BFET_Soft_OffDiag}
		\bigr[\Gamma^{\mathcal{Q}\ne0;soft}_{(t)}\bigr]^{a}_{off-diag}  =& g^{2} \delta_{ab} \sum_{\substack{c,d,e,f,g=1 \\ (d\ne e)}}^{N_{c}}
		\bigr(t^{de}_{ac}t^{fg}_{cb}\bigr) \int_q \int d\omega \;
		\delta(\omega-\vec{v}_{1}\cdot\vec{q}\;) \mathcal{N}(\omega,\mathcal{Q}^{de})
		\biggr\{ \rho_{L,off-diag}^{de,fg} + \vec{v}_{1}^{\;2}
		\bigr[ 1-(\widehat{v}_{1}\cdot\widehat{q})^{2} \bigr]\rho_{T,off-diag}^{de,fg} \biggr\}.
	\end{aligned}
\end{equation}
The color-averaged scattering rate [Eq.~(\ref{eq:Gamma_ColorAvg_BFET})] can be obtained as
\begin{equation}
	\begin{aligned} \label{eq:Gamma_ColorAvg_BFET_Soft_OffDiag}
		\bigr[\Gamma^{\mathcal{Q}\ne0;soft}_{(t)}(E_{1},T) \bigr]_{off-diag} =& \frac{g^{2}}{2N_{c}} \int_q \int d\omega \;
		\delta(\omega-\vec{v}_{1}\cdot\vec{q}\;)
		\sum_{\substack{d,e=1 \\ (d\ne e)}}^{N_{c}} \mathcal{N}(\omega,\mathcal{Q}^{de})\biggr\{ \rho_{L,off-diag}^{de,ed} + \vec{v}_{1}^{\;2}
		\bigr[ 1-(\widehat{v}_{1}\cdot\widehat{q})^{2} \bigr]\rho_{T,off-diag}^{de,ed} \biggr\},
	\end{aligned}
\end{equation}
with the color factor in Eq.~(\ref{eq:Gamma_BFET_Soft_OffDiag})
\begin{equation}
	\begin{aligned} \label{eq:BFET_ColorFact_OffDiag}
		\sum_{a,c,d,e,f,g=1}^{N_{c}} (\delta_{ab} t^{de}_{ac}t^{fg}_{cb} \delta^{dg}\delta^{ef}) = Tr(t^{de}t^{ed})
		&= \sum_{d,e=1}^{N_{c}} \frac{1}{2}(1-\frac{\delta^{de}}{N_{c}})
		\stackrel{d\ne e}{=} \frac{1}{2} \sum_{\substack{d,e=1 \\ (d\ne e)}}^{N_{c}}.
	\end{aligned}
\end{equation}

A similar procedure can be applied to the diagonal gluons,
yielding the color-averaged scattering rate given by
\begin{equation}
	\begin{aligned} \label{eq:Gamma_ColorAvg_BFET_Soft_Diag}
		\bigr[\Gamma^{\mathcal{Q}\ne0;soft}_{(t)}\bigr]_{diag}(E_{1},T)  &= \frac{g^{2}}{2N_{c}} \int_q \int d\omega \;
		\delta(\omega-\vec{v}_{1}\cdot\vec{q}\;)
		\sum_{h=1}^{N_{c}-1} \bar{n}_{B}(\omega) 
		\biggr\{ \rho^{h}_{L,diag} + \vec{v}_{1}^{\;2}
		\bigr[ 1-(\widehat{v}_{1}\cdot\widehat{q})^{2} \bigr]\rho^{h}_{T,diag} \biggr\},
	\end{aligned}
\end{equation}
with the color factor as in Eq.~(\ref{eq:BFET_Trace1})
\begin{equation}
	\begin{aligned} \label{eq:BFET_ColorFact_Diag}
		\sum_{a,c=1}^{N_{c}} \sum_{\substack{d,e,f,g=1 \\ (d=e,f=g)}}^{N_{c}-1} (\delta_{ab} t^{de}_{ac}t^{fg}_{cb})
		&= \sum_{d,f=1}^{N_{c}-1} \biggr[\sum_{a,c=1}^{N_{c}} (t^{dd}_{ac}t^{ff}_{ca})\biggr]
		=  \frac{1}{2} \sum_{d,f=1}^{N_{c}-1} \mathcal{P}^{dd,ff}.
	\end{aligned}
\end{equation}
The corresponding transverse and longitudinal spectral functions read
\begin{subequations}
	\begin{align}
		\rho^{h}_{T,diag}(\omega,q) =& \frac{\pi \omega (\mathcal{M}^{2}_{D})^{h}_{diag}}{2q^{3}} (q^{2}-\omega^{2})
		\Biggr\{ \biggr[ q^{2}-\omega^{2} + \frac{\omega^{2} (\mathcal{M}^{2}_{D})^{h}_{diag}}{2q^{2}}
		\bigr(1+\frac{q^{2}-\omega^{2}}{2\omega q} ln\frac{q+\omega}{q-\omega}\bigr) \biggr]^{2} \nonumber \\
		& + \biggr[ \frac{\pi \omega (\mathcal{M}^{2}_{D})^{h}_{diag}}{4q^{3}} (q^{2}-\omega^{2}) \biggr]^{2} \Biggr\}^{-1}, \label{eq:BFET_GammaRhoT_Soft_Diag} \\
		\rho^{h}_{L,diag}(\omega,q) =& \frac{\pi \omega (\mathcal{M}^{2}_{D})^{h}_{diag}}{q}
		\Biggr\{ \biggr[ q^{2}+ (\mathcal{M}^{2}_{D})^{h}_{diag}\bigr(1-\frac{\omega}{2q} ln\frac{q+\omega}{q-\omega}\bigr) \biggr]^{2} + 
		\biggr( \frac{\pi \omega (\mathcal{M}^{2}_{D})^{h}_{diag}}{2q} \biggr)^{2} \Biggr\}^{-1}. \label{eq:BFET_GammaRhoL_Soft_Diag}
	\end{align}
\end{subequations}

Combining the contributions from the off-diagonal [Eq.~(\ref{eq:Gamma_ColorAvg_BFET_Soft_OffDiag})]
and diagonal gluons [Eq.~(\ref{eq:Gamma_ColorAvg_BFET_Soft_Diag})],
we arrive at Eq.~(\ref{eq:Gamma_ColorAvg_BFET_Soft}).


\section{Calculation of the color factors in the double line basis} \label{appendix:ColorFactor_BFET}
\setcounter{equation}{0}
\renewcommand\theequation{B\arabic{equation}}

In this Appendix, we calculate the color factors for heavy quark
scattering off gluons in $t$-, $s$- and $u$ channels for $\mathcal{Q}\ne0$ in the double line basis,
which will be compared with the corresponding results for $\mathcal{Q}=0$.

\begin{figure}[!htbp]
	\centering
	\includegraphics[width=.32\textwidth]{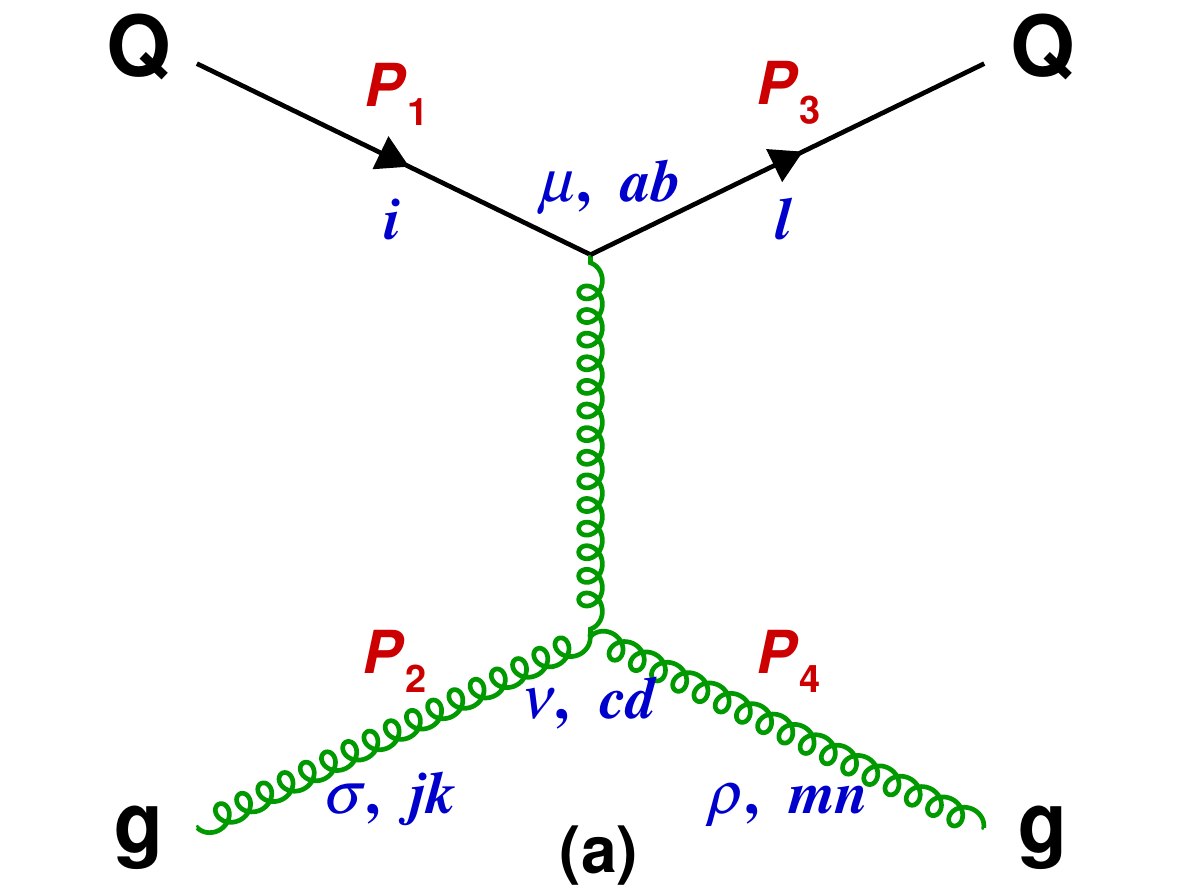}
	\includegraphics[width=.32\textwidth]{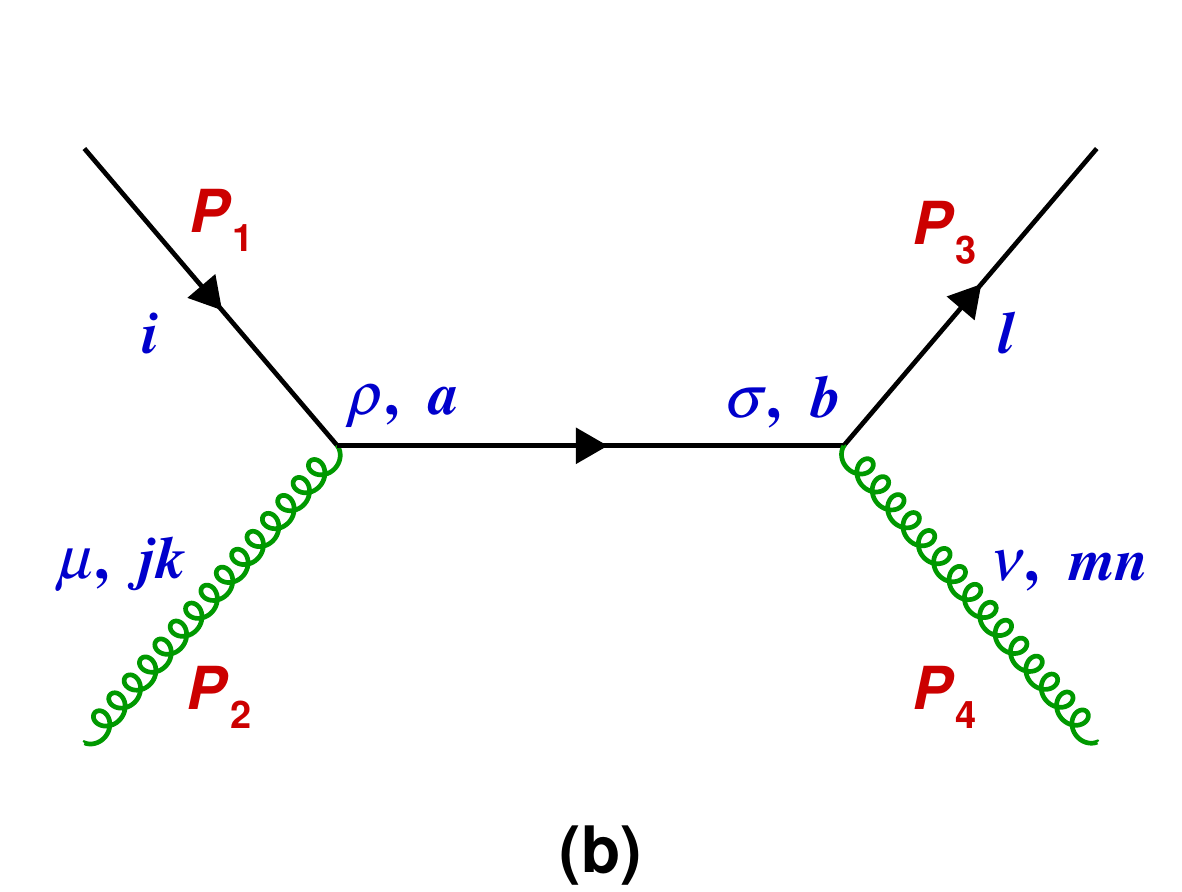}
	\includegraphics[width=.32\textwidth]{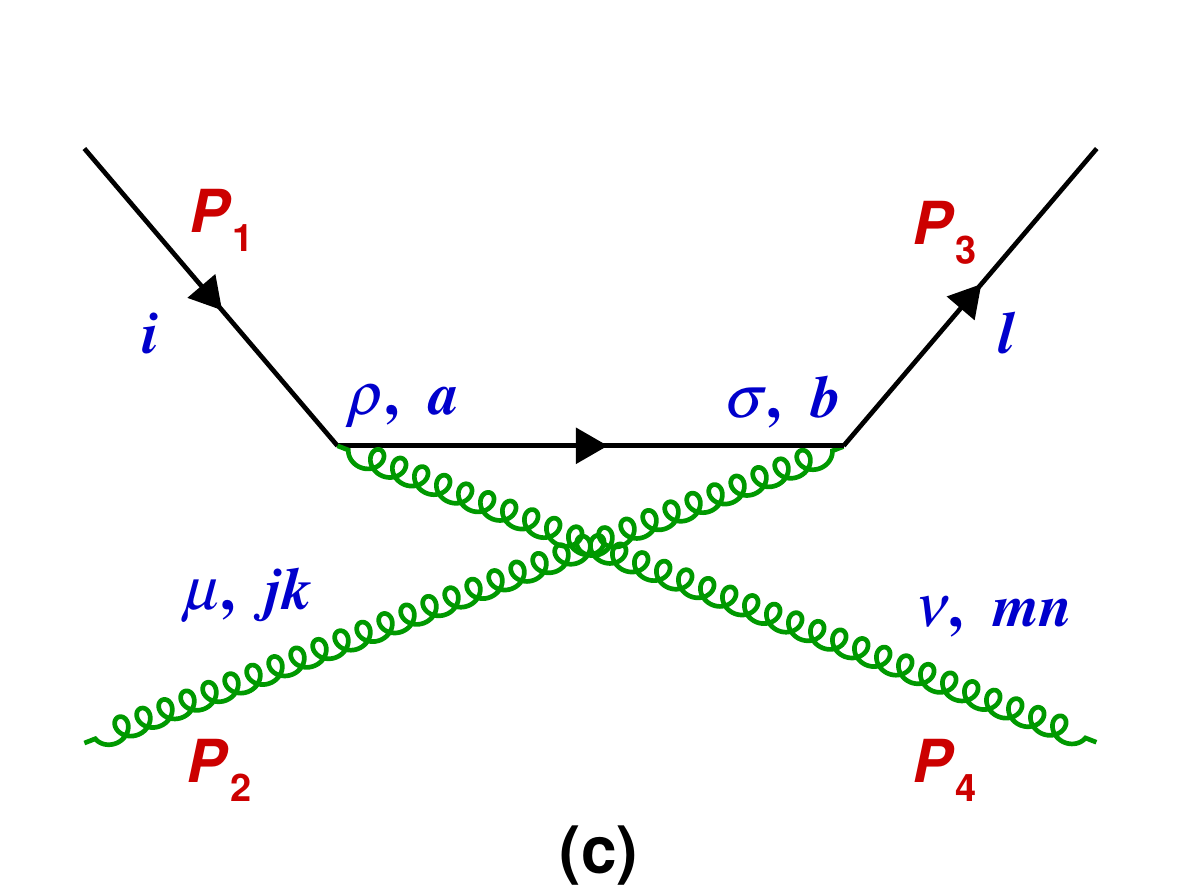}
	\caption{Same as Fig.~\ref{fig:TwoBody_Diag} but with the
		background field effective theory in the double line basis.}
	\label{fig:TwoBody_Diag_BFET}
\end{figure}
Figure~\ref{fig:TwoBody_Diag_BFET} shows 
the Feynman diagrams for tree-level quark-gluon scattering
in different channels in the double line basis.
The color factor for the scattering in the $t$ channel (panel a in Fig.~\ref{fig:TwoBody_Diag_BFET}) reads
\begin{equation}
	\begin{aligned} \label{eq:ColorFactor_QgTchannel_BFET}
		\biggr[C_{F}\biggr]^{\mathcal{Q}\ne0}_{(t)} =& -\frac{1}{N_{c}}\sum^{N_{c}}_{i=1}
		\frac{1}{N_{c}^{2}-1}\sum^{N_{c}}_{j,k=1}
		\biggr[ \sum^{N_{c}}_{l=1} \sum^{N_{c}}_{m,n=1} \bigr( t^{ab}_{li} \mathcal{P}_{ab,cd} f^{cd,mn,jk} \bigr)
		\cdot \bigr( t^{ef}_{il} \mathcal{P}_{ef,gh} f^{gh,nm,kj} \bigr) \biggr]
		= \frac{1}{16} \sum_{j,k=1}^{3} \left(1-\frac{\delta^{jk}}{3}\right),
	\end{aligned}
\end{equation}
where, $f^{cd,mn,jk}=i(\delta^{cn}\delta^{mk}\delta^{jm} - \delta^{ck}\delta^{md}\delta^{jn})/\sqrt{2}$
is the structure constant for $SU(N_{c}=3)$ in the double line basis.
The color factor for the similar process but for $\mathcal{Q}=0$ (panel a in Fig.~\ref{fig:TwoBody_Diag}) is given by
\begin{equation}
	\begin{aligned} \label{eq:ColorFactor_QgTchannel_Pert}
		\biggr[C_{F}\biggr]^{\mathcal{Q}=0}_{(t)} =& \frac{1}{N_{c}}\sum^{N_{c}}_{i=1}
		\frac{1}{N_{c}^{2}-1}\sum^{N_{c}^{2}-1}_{j=1}
		\biggr[ \sum^{N_{c}}_{l=1} \sum^{N_{c}^{2}-1}_{m} \bigr( T^{a}_{li} \delta_{ac} f^{cmj} \bigr)
		\cdot \bigr( T^{e}_{il} \delta_{eg} f^{gmj} \bigr) \biggr]                                                                                                    
		= \frac{1}{2},
	\end{aligned}
\end{equation}
where, $T^{a}_{li}$ and $f^{cmj}$ are the corresponding generator and structure constant, respectively.
Thus, the scattering amplitude for quark-gluon scattering in $t$ channel for $\mathcal{Q}\ne0$ reads
\begin{equation}
	\begin{aligned}\label{eq:MforQ_t}
		\overline{|\mathcal{M}^{2}|}^{\mathcal{Q}\ne0}_{(t)} = \biggr[C_{F}\biggr]^{\mathcal{Q}\ne0}_{(t)} \biggr/
		\biggr[C_{F}\biggr]^{\mathcal{Q}=0}_{(t)} \cdot \overline{|\mathcal{M}^{2}|}^{\mathcal{Q}=0}_{(t)}
		= \frac{1}{8}\sum_{j,k=1}^{3}\left(1-\frac{1}{3}\delta^{jk}\right) \overline{|\mathcal{M}^{2}|}^{\mathcal{Q}=0}_{(t)},
	\end{aligned}
\end{equation}
where, $\overline{|\mathcal{M}^{2}|}^{\mathcal{Q}=0}_{(t)}$ is the relevant scattering amplitude
obtained in the absence of the background field $\mathcal{Q}=0$~\cite{Peng:2024zvf, Li:2021nim}.

The color factor for the scattering in the $s$ channel (panel b in Fig.~\ref{fig:TwoBody_Diag_BFET}) reads
\begin{equation}
	\begin{aligned} \label{eq:ColorFactor_QgSchannel_BFET}
		\biggr[C_{F}\biggr]^{\mathcal{Q}\ne0}_{(s)} =& \frac{1}{N_{c}}\sum^{N_{c}}_{i=1}
		\frac{1}{N_{c}^{2}-1}\sum^{N_{c}}_{j,k=1}
		\biggr[ \sum^{N_{c}}_{l=1} \sum^{N_{c}}_{m,n=1} \bigr( t^{mn}_{lb}  t^{jk}_{bi})
		\cdot \bigr( t^{jk}_{ia} t^{nm}_{al} \bigr) \biggr]
		= \frac{1}{36} \sum_{j,k=1}^{3} \left(1-\frac{\delta^{jk}}{3}\right).
	\end{aligned}
\end{equation}
The color factor for the similar process but for $\mathcal{Q}=0$ (panel b in Fig.~\ref{fig:TwoBody_Diag}) is given by
\begin{equation}
	\begin{aligned} \label{eq:ColorFactor_QgSchannel_Pert}
		\biggr[C_{F}\biggr]^{\mathcal{Q}=0}_{(s)} =& \frac{1}{N_{c}}\sum^{N_{c}}_{i=1}
		\frac{1}{N_{c}^{2}-1}\sum^{N_{c}^{2}-1}_{j=1}
		\biggr[ \sum^{N_{c}}_{l=1} \sum^{N_{c}^{2}-1}_{m} \bigr( T^{m}_{lb} T^{j}_{bi} \bigr)
		\cdot \bigr( T^{j}_{ia} T^{m}_{al} \bigr) \biggr]                                                                                                    
		= \frac{2}{9}.
	\end{aligned}
\end{equation}
It appears that,
very similar to the results in the $t$ channel [Eq.~(\ref{eq:MforQ_t})],
the quark-gluon scattering amplitude in the $s$ channel is expressed as,
\begin{equation}
	\begin{aligned}\label{eq:MforQ_s}
		\overline{|\mathcal{M}^{2}|}^{\mathcal{Q}\ne0}_{(s)} = \biggr[C_{F}\biggr]^{\mathcal{Q}\ne0}_{(s)} \biggr/
		\biggr[C_{F}\biggr]^{\mathcal{Q}=0}_{(s)} \cdot \overline{|\mathcal{M}^{2}|}^{\mathcal{Q}=0}_{(s)}
		= \frac{1}{8}\sum_{j,k=1}^{3}\left(1-\frac{1}{3}\delta^{jk}\right) \overline{|\mathcal{M}^{2}|}^{\mathcal{Q}=0}_{(s)},
	\end{aligned}
\end{equation}
where, $\overline{|\mathcal{M}^{2}|}^{\mathcal{Q}=0}_{(s)}$ is the relevant scattering amplitude
obtained in the absence of the background field $\mathcal{Q}=0$.
Similar results can be obtained in the $u$ channels,
as well as the interference terms.

The color factor in the $u$ channel for $\mathcal{Q}\ne0$ (panel c in Fig.~\ref{fig:TwoBody_Diag_BFET}) is given by
\begin{equation}
	\begin{aligned} \label{eq:ColorFactor_QgUchannel_BFET}
		\biggr[C_{F}\biggr]^{\mathcal{Q}\ne0}_{(u)} =& \frac{1}{N_{c}}\sum^{N_{c}}_{i=1}                                                                             
		\frac{1}{N_{c}^{2}-1}\sum^{N_{c}}_{j,k=1}
		\biggr[ \sum^{N_{c}}_{l=1} \sum^{N_{c}}_{m,n=1} \bigr( t^{jk}_{lb}  t^{mn}_{bi})
		\cdot \bigr( t^{nm}_{ia} t^{kj}_{al} \bigr) \biggr]
		= \frac{1}{36} \sum_{j,k=1}^{3} \left(1-\frac{\delta^{jk}}{3}\right),
	\end{aligned}
\end{equation}
and the result for $\mathcal{Q}=0$ (panel c in Fig.~\ref{fig:TwoBody_Diag}) is expressed as
\begin{equation}
	\begin{aligned} \label{eq:ColorFactor_QgUchannel_Pert}
		\biggr[C_{F}\biggr]^{\mathcal{Q}=0}_{(u)} =& \frac{1}{N_{c}}\sum^{N_{c}}_{i=1}
		\frac{1}{N_{c}^{2}-1}\sum^{N_{c}^{2}-1}_{j=1}
		\biggr[ \sum^{N_{c}}_{l=1} \sum^{N_{c}^{2}-1}_{m} \bigr( T^{j}_{lb} T^{m}_{bi} \bigr)
		\cdot \bigr( T^{m}_{ia} T^{j}_{al} \bigr) \biggr]
		= \frac{2}{9}.
	\end{aligned}
\end{equation}

Concerning the interference terms, we just list the final results below:
\begin{equation}
	\begin{aligned} \label{eq:ColorFactor_QgTSstarchannel}
		\biggr[C_{F}\biggr]^{\mathcal{Q}\ne0}_{(ts^{\ast})} = -\frac{i}{32} \sum_{j,k=1}^{3} \left(1-\frac{\delta^{jk}}{3}\right), \qquad
		\biggr[C_{F}\biggr]^{\mathcal{Q}=0}_{(ts^{\ast})} = -\frac{i}{4},
	\end{aligned}
\end{equation}
\begin{equation}
	\begin{aligned} \label{eq:ColorFactor_QgTUstarchannel}
		\biggr[C_{F}\biggr]^{\mathcal{Q}\ne0}_{(tu^{\ast})} = \frac{i}{32} \sum_{j,k=1}^{3} \left(1-\frac{\delta^{jk}}{3}\right), \qquad
		\biggr[C_{F}\biggr]^{\mathcal{Q}=0}_{(tu^{\ast})} = \frac{i}{4},
	\end{aligned}
\end{equation}
\begin{equation}
	\begin{aligned} \label{eq:ColorFactor_QgSUstarchannel}
		\biggr[C_{F}\biggr]^{\mathcal{Q}\ne0}_{(su^{\ast})} = -\frac{1}{288} \sum_{j,k=1}^{3} \left(1-\frac{\delta^{jk}}{3}\right), \qquad
		\biggr[C_{F}\biggr]^{\mathcal{Q}=0}_{(su^{\ast})} = -\frac{1}{36}.
	\end{aligned}
\end{equation}



\bibliography{BFET}
\end{document}